\documentclass[11pt,a4paper,reqno]{article}
\pdfoutput=1
\def\arXiv{1}

\usepackage[margin=3.4cm]{geometry}

\usepackage{authblk}

\usepackage{amsmath}
\usepackage{amsfonts}
\usepackage{amsthm}
\usepackage{amssymb}
\numberwithin{equation}{section}

\usepackage{setspace}

\usepackage{titlesec}
\titleformat{\section}{\large\bfseries}{\thesection}{1.5ex}{}
\titleformat{\subsection}{\normalsize\bfseries}{\thesubsection}{1.5ex}{}

\usepackage{graphicx}
\usepackage{booktabs}
\usepackage[suffix=]{epstopdf}
\usepackage[font=footnotesize,labelfont=bf]{caption}
\usepackage[colorlinks]{hyperref}
\hypersetup{citecolor=blue,urlcolor=blue,linkcolor=blue}
\newtheorem{theorem}{Theorem}
\newtheorem{proposition}{Proposition}
\newtheorem*{conjecture*}{Conjecture}
\theoremstyle{definition}

\newtheorem{example}{Example}
\theoremstyle{remark}
\newtheorem{remark}{Remark}
\usepackage{mathrsfs} 
\usepackage{mathtools} 
\usepackage{booktabs}
\usepackage{tabularx}
\newcolumntype{L}{>{\raggedright\arraybackslash}X}

\usepackage{bbding}

\usepackage[shortlabels]{enumitem}

\usepackage[sort&compress,numbers]{natbib}
\usepackage{cleveref}
\crefname{equation}{}{}
\crefname{enumi}{}{}
\crefname{theorem}{Theorem}{Theorems}
\crefname{corollary}{Corollary}{Corollaries}
\crefname{example}{Example}{Examples}
\crefname{lemma}{Lemma}{Lemmas}
\crefname{proposition}{Proposition}{Propositions}
\crefname{figure}{Figure}{Figures}
\crefname{table}{Table}{Tables}
\crefformat{section}{\S#2#1#3}
\crefmultiformat{section}{\S#2#1#3}{ and~\S#2#1#3}{, \S#2#1#3}{, and~\S#2#1#3}
\crefformat{subsection}{\S#2#1#3}
\crefmultiformat{subsection}{\S#2#1#3}{ and~\S#2#1#3}{, \S#2#1#3}{, and~\S#2#1#3}
\crefname{appendix}{appendix}{appendices}
\Crefname{equation}{}{}
\Crefname{enumi}{}{}
\Crefname{theorem}{Theorem}{Theorems}
\Crefname{corollary}{Corollary}{Corollaries}
\Crefname{example}{Example}{Examples}
\Crefname{lemma}{Lemma}{Lemma}
\Crefname{proposition}{Proposition}{Proposition}
\Crefname{figure}{Figure}{Figures}
\Crefname{table}{Table}{Tables}
\Crefformat{section}{Section~#2#1#3}
\Crefmultiformat{section}{Sections~#2#1#3}{ and~#2#1#3}{, #2#1#3}{, and~#2#1#3}
\Crefformat{subsection}{Section~#2#1#3}
\Crefmultiformat{subsection}{Sections~#2#1#3}{ and~#2#1#3}{, #2#1#3}{, and~#2#1#3}
\Crefname{appendix}{Appendix}{Appendices}

\newcommand{\norm}[2]{\left\| #1 \right\|_{#2}}
\newcommand{\ip}[2]{\langle #1 , #2 \rangle}
\newcommand{\bigip}[2]{\left\langle #1 , #2 \right\rangle}
\newcommand{\abs}[1]{\left\vert #1 \right\vert}

\newcommand{\dVolume}{{\rm d}\vec{x}}

\renewcommand{\vec}[1]{\mathbf{#1}}

\renewcommand{\theta}{\vartheta}
\renewcommand{\phi}{\varphi}
\newcommand{\vecphi}{\boldsymbol{\phi}}
\newcommand{\vecxi}{\boldsymbol{\xi}}
\newcommand{\veceta}{\boldsymbol{\eta}}
\newcommand{\veclambda}{\boldsymbol{\lambda}}
\newcommand{\maxAverage}{\overline{\Phi}^*}
\newcommand{\backgroundField}{\vec{U}}
\newcommand{\LieDer}{\mathcal{L}}

\newcommand{\opA}{\mathscr{A}}
\newcommand{\opB}{\mathscr{B}}
\newcommand{\opC}{\mathscr{C}}

\newcommand{\opH}{\mathscr{H}}
\newcommand{\opL}{\mathscr{L}}

\newcommand{\opK}{\mathscr{K}}
\newcommand{\opF}{\mathscr{F}}
\newcommand{\quadratic}{q}
\newcommand{\linear}{\ell}
\newcommand{\opP}{\mathscr{P}}

\DeclareMathOperator{\linspan}{span}

\DeclareMathOperator{\rank}{rank}

\newcommand{\group}{\mathbb{G}}
\newcommand{\GroupAction}{\mathscr{G}}

\newcommand{\bR}{\mathbb{R}}        
\newcommand{\uSpace}{\mathbb{U}}    
\newcommand{\vSpace}{\mathbb{V}}    
\newcommand{\wSpace}{\mathbb{W}}    
\newcommand{\pSpace}{\mathbb{P}}    
\newcommand{\absSet}{\mathbb{A}}    
\newcommand{\energyPreservingSpace}{\mathbb{M}}

\newcommand{\ddt}[1]{\frac{{\rm d} #1}{{\rm d}t}}

\DeclareMathOperator{\Rey}{Re}
\DeclareMathOperator{\Ra}{Ra}

\DeclareMathOperator{\Pran}{Pr}

\newcommand{\tr}{{\scriptscriptstyle\top}}

\author{Giovanni Fantuzzi\thanks{Corresponding author: \href{mailto:giovanni.fantuzzi10@imperial.ac.uk}{giovanni.fantuzzi10@imperial.ac.uk}}}
\author{Ali Arslan}
\author{Andrew Wynn} 
\affil{\small Department of Aeronautics, Imperial College London, London, SW7 2AZ, UK}
\date{\today}

\title{\Large\bf The background method: Theory and computations}

\begin{document}


\maketitle

\vspace*{-20pt}
\begin{abstract}\noindent
    The background method is a widely used technique to bound mean properties of turbulent flows rigorously. This work reviews recent advances in the theoretical formulation and numerical implementation of the method. First, we describe how the background method can be formulated systematically within a broader ``auxiliary function'' framework for bounding mean quantities, and explain how symmetries of the flow and constraints such as maximum principles can be exploited. All ideas are presented in a general setting and are illustrated on Rayleigh--B\'enard convection between stress-free isothermal plates. Second, we review a semidefinite programming approach and a timestepping approach to optimizing bounds computationally, revealing that they are related to each other through convex duality and low-rank matrix factorization. Open questions and promising directions for further numerical analysis of the background method are also outlined.
\end{abstract}

{\small\noindent
\textbf{MSC 2020 subject classification:} 76M30, 76F25, 90C22\\[1ex]
\textbf{Keywords:} Background method, Bounds, Auxiliary functions, Variational methods,\\Semidefinite programming
}

\section{Introduction}\label{s:intro}
Making quantitative predictions for key properties of incompressible turbulent flows, such as the mean energy dissipation in a pipe or the average amount of heat transported by natural convection, is a fundamental problem in fluid mechanics. A particular challenge is to determine the functional dependence of these mean quantities on relevant nondimensional parameters, such as the Reynolds number, directly from the equations governing the flow and without introducing physically reasonable, but unproven, assumptions.

One way to study mean quantities rigorously, pioneered by Malkus, Howard and Busse~\cite{Malkus1954,Howard1963,Busse1969,Busse1970,Howard1972,Busse1979}, is to bound their values using variational techniques. The idea is simple: rather than optimizing the mean quantity of interest over solutions to the flow's governing equations, one optimizes it over a larger set of incompressible flow fields that satisfy only integral constraints in the form of energy and flux balances, which are weaker but more tractable. The latter optimum is often still hard to evaluate, but conservative one-sided estimates (lower bounds for minima and upper bounds for maxima) can be obtained with the background method of Doering\&Constantin~\cite{Doering1992,Doering1994,Constantin1995a,Doering1996}.

To illustrate the method, suppose we seek bounds on the mean energy dissipation for the flow of an incompressible fluid in a domain $\Omega$ with no-slip boundaries driven by an externally applied force. The velocity $\vec{u}$ of the fluid is governed by the nondimensional Navier--Stokes equations
\begin{subequations}
\label{e:intro-ns}
\begin{gather}
\label{e:intro-ns-momentum}
\partial_t \vec{u} = \Rey^{-1} \Delta \vec{u} - \vec{u} \cdot \nabla \vec{u} - \nabla p + \vec{f},\\
\nabla \cdot \vec{u} = 0,
\end{gather}
\end{subequations}
where $\Rey$ is the Reynolds number, $p$ is the pressure, and $\vec{f}$ is the nondimensional external forcing. The mean energy dissipation is
\begin{equation}
\overline{ \Phi } = \overline{ \int_\Omega \abs{\nabla \vec{u}}^2 \dVolume },
\end{equation}
where we use overbars to indicate averaging over infinite time.
The background method begins by considering the energy $\frac12 \int_{\Omega} \abs{\vec{u} - \vec{U}}^2 \dVolume$ of perturbations from a \textit{background field} $\vec{U}$, which is incompressible and satisfies no-slip boundary conditions, but can otherwise be chosen arbitrarily. Assuming that this perturbation energy remains uniformly bounded in time, as is reasonable to expect for a broad class of force vectors $\vec{f}$, its time derivative has zero infinite-time average:
\begin{align}
0 
&= \overline{ \ddt{} \frac12 \int_{\Omega} \abs{\vec{u} - \vec{U}}^2 \dVolume } \nonumber \\
&= \overline{ \int_{\Omega} \left( \vec{u} - \vec{U} \right) \cdot \left( \Rey^{-1} \Delta \vec{u} - \vec{u} \cdot \nabla \vec{u} - \nabla p + \vec{f}\right) \dVolume} \nonumber \\
&= \overline{  \int_{\Omega} \left(
-\Rey^{-1} \abs{\nabla \vec{u}}^2 + \Rey^{-1} \nabla \vec{U} \cdot \nabla \vec{u} 
- \vec{u} \cdot \nabla \vec{U} \cdot \vec{u} + \vec{f}\cdot\vec{u} - \vec{f}\cdot\vec{U} \right) \dVolume }.
\end{align}
Here, the second line follows from differentiation in time using the momentum equation~\cref{e:intro-ns-momentum}, while the last line is obtained after integration by parts using the no-slip boundary conditions and the incompressibility of $\vec{u}$ and $\vec{U}$. This identity enables us to express the mean energy dissipation as
\begin{equation}
\label{e:intro-dissipation}
\overline{ \Phi }
=
\overline{  \int_{\Omega} \left(
(1-\tfrac{\lambda}{\Rey}) \abs{\nabla \vec{u}}^2 + \tfrac{\lambda}{\Rey} \nabla \vec{U} \cdot \nabla \vec{u} 
- \lambda \vec{u} \cdot \nabla \vec{U} \cdot \vec{u} + \lambda \vec{f}\cdot\vec{u} - \lambda \vec{f}\cdot\vec{U} \right) \dVolume },
\end{equation}
where $\lambda$ is an arbitrary scalar often called the \textit{balance parameter}. Suppose now that, for a given value of $\Rey$, the background field $\vec{U}$ and the balance parameter $\lambda$ can be chosen such that
\begin{equation}\label{e:intro-bounding-ineq}
\int_{\Omega} \left(
(1-\tfrac{\lambda}{\Rey}) \abs{\nabla \vec{u}}^2 + \tfrac{\lambda}{\Rey} \nabla \vec{U} \cdot \nabla \vec{u} 
- \lambda \vec{u} \cdot \nabla \vec{U} \cdot \vec{u} + \lambda \vec{f}\cdot\vec{u} - \lambda \vec{f}\cdot\vec{U} \right) \leq \gamma
\end{equation}
for some constant $\gamma$ and all incompressible vector fields $\vec{u}$ satisfying the no-slip boundary conditions. Then, the inequality holds also for all possible solutions of the Navier--Stokes equations (\ref{e:intro-ns}a,b) and identity~\cref{e:intro-dissipation} implies the upper bound $\overline{\Phi} \leq \gamma$ at the given Reynolds number. To obtain a lower bound $\overline{\Phi} \geq \gamma$, it suffices to reverse inequality~\cref{e:intro-bounding-ineq}. One is of course especially interested in obtaining bounds $\gamma=\gamma(\Rey)$ for all or a range of Reynolds numbers, so as to understand how the mean energy dissipation (the mean quantity of interest) scales with $\Rey$ (the control parameter).  As demonstrated in \cref{ss:general-background-method}, these ideas can be extended to a very broad class of flows by considering generalized perturbation energies and generalized background fields.

In the above example and in general, optimizing the background field and the balance parameter to produce the best possible bound is a dual problem (in the sense of convex duality) to the Malkus--Howard--Busse approach described above~\cite{Kerswell1998,Kerswell1999,Kerswell2001} 
and is generally difficult. Suboptimal background fields, however, can usually be constructed using only elementary calculus and functional inequalities, and often yield useful bounds. For these reasons, the background method has enjoyed tremendous success since its introduction in the 1990s (see \cref{table:list_of_addressed_problems} for a nonexhaustive list of flows to which it has been applied) and, to this date, it remains one of the key tools for rigorous flow analysis.

\begin{table}
\centering
\caption{A nonexhaustive summary of flows to which the background method has been applied.}
\label{table:list_of_addressed_problems}
{
\ifx\arXiv\undefined
\footnotesize
\else
\if\arXiv1
\footnotesize
\fi
\fi
\begin{tabularx}{\linewidth}{XlccX} 
	\toprule
	&Flow & Analysis & Computations &\\
	\midrule
	&Rayleigh--B\'enard convection  & 
	\cite{Doering1996,Constantin1996a,Kerswell1997,Constantin1999,Doering2001,Kerswell2001,Otero2002,Yan2004,Plasting2005,Ierley2006,Doering2006,Wittenberg2010,Whitehead2011prl,Otto2011,Whitehead2012,Whitehead2014,Goluskin2016d,Nobili2017,Fantuzzi2018,Pachev2020,Christopher2021} &
	\cite{Doering1997,Plasting2005,Ierley2006,Wittenberg2010a,Wen2015a,Tilgner2017,Fantuzzi2018,Tilgner2019,Pachev2020,Ding2020}, \cite{Plasting2003}$^\dagger$\\
	&B\'enard--Marangoni convection & \cite{Hagstrom2010,Fantuzzi2020jfm} & \cite{Fantuzzi2018a} \\ 
	&Porous-media convection & \cite{Doering1998} & \cite{Otero2004,Wen2012,Wen2013,Wen2018} \\
	&Internally heated convection & \cite{Lu2004,Whitehead2011,Whitehead2012,Goluskin2015a,Arslan2021,Arslan2021a,KumarArslan2021} & \cite{Arslan2021,Arslan2021a} \\
	&Double-diffusive convection & \cite{Balmforth2006} & none \\
	&Horizontal convection & \cite{siggers2004} & none \\
	&Parallel shear flows & \cite{Doering1992,Doering1994,Marchioro1994,Kerswell1997,Nicodemus1997,Nicodemus1998,Kerswell1998,Doering2000,Kerswell2002,Hagstrom2014} & \cite{Nicodemus1997a,Nicodemus1998a,Plasting2003,Tang2004,Fantuzzi2016PRE,Lee2019}\\
	&Taylor--Couette flow & \cite{Constantin1994,Gallet2010,Ding2019} & \cite{Ding2019}\\
	&Pressure-driven channel flow & \cite{Constantin1995a,Kumar2020a} & none\\
	&Precessing flow & \cite{Kerswell1996} & none \\
	&Flows in unbounded domains & \cite{Kumar2020,Tilgner2021} & none \\
	\bottomrule
\end{tabularx}}
\\
\raggedright\scriptsize
$^\dagger$Computations for Couette flow in~\cite{Plasting2003} imply optimal bounds for Rayleigh--B\'enard convection~\cite[\S4]{Kerswell2001}.
\end{table}

This work reviews recent advances in the theoretical formulation and computational implementation of the background method. The latter are of particular interest because, even though no algorithm can produce rigorous bounds for all possible values of a flow's governing parameters, numerical approximations of the best bound available to the background method can reveal whether suboptimal bounds proven analytically exhibit optimal parameter dependence. If they do not, moreover, computations can guide improved analysis. This was recently demonstrated for B\'enard--Marangoni convection at infinite Prandtl number~\cite{Fantuzzi2018a,Fantuzzi2020jfm} and for internally heated convection~\cite{Arslan2021,KumarArslan2021}.

On the theoretical side, we show how to formulate the background method systematically within a more general framework for bounding infinite-time averages, which is based on auxiliary functions~\cite{Chernyshenko2014a,Fantuzzi2016b,Goluskin2019} and is known to be sharp for well-posed ordinary and partial differential equations~\cite{Tobasco2018,Rosa2020}. This interpretation of the background method, discussed in~\cite{Chernyshenko2014a,Chernyshenko2017,Fantuzzi2018,Goluskin2019,Arslan2021,Arslan2021a} and already used in the example given above, will be described in a general setting in \cref{ss:general-background-method}. Its main advantages compared to the original formulation of the method by Doering \& Constantin~\cite{Doering1992,Doering1994,Constantin1995a,Doering1996} 
are that (i) it allows for the consideration of generalized perturbation energies and generalized background fields, (ii) it can be used easily to bound mean quantities not equivalent to the dissipation rate, and (iii) it always reduces the search for a bound to a \textit{convex} variational principle. Indeed, observe that inequality~\cref{e:intro-bounding-ineq} is jointly convex in the balance parameter $\lambda$, the scaled background field $\lambda \vec{U}$, and the bound $\gamma$. We also explain how this convex variational principle can be simplified using symmetries and, sometimes, improved by incorporating constraints such as maximum principles.

On the computational side (\cref{s:implementation}), we first review two recent approaches to optimizing background fields---one based on semidefinite programming~\cite{Fantuzzi2015,Fantuzzi2016PRE,Fantuzzi2018a,Tilgner2017} and one based on timestepping~\cite{Wen2013,Wen2015a}---that stand out from other strategies employed in the literature (cf.~\cref{table:list_of_addressed_problems}) for their generality, robustness, and simplicity.
We then reveal a previously unrecognised connection between these two approaches, which opens new avenues for the numerical analysis of the background method and may enable the efficient optimization of bounds for complex flows.   

To highlight the broad applicability of the background method (including its numerical implementation strategies) and separate its key ingredients from specific aspects pertaining to particular examples, we work in an abstract setting similar to that in~\cite{Constantin1995}, which generalizes the basic Navier--Stokes equations in~\cref{e:intro-ns} but preserves their key properties (see \cref{s:governing-equations}). To minimize the technicalities, however, we omit a rigorous functional-analytic setup. The abstract ideas are illustrated in the context of Rayleigh--B\'enard convection between stress-free isothermal plates.

\section{Governing equations in abstract form}\label{s:governing-equations}

We consider an incompressible fluid in a domain $\Omega \subset \bR^d$ (usually, $d=2$ or~$3$), whose state at time $t$ is described by a pressure field $p(t):\Omega \to \bR$ and a vector field $\vec{u}(t):\Omega \to \bR^n$. The components of $\vec{u}$ represent $n$ physical variables of interest, such as velocity, temperature, and vorticity. At each time $t$, the state $\vec{u}(t)$ and pressure $p(t)$ are functions of the spatial coordinate $\vec{x} \in \Omega$, and we assume that:
\begin{enumerate}[({A}1), leftmargin=*, labelsep=1em, widest=9, topsep=1.5ex, itemsep=0ex]
\item\label{ass:subspaces} $\vec{u}(t)$ belongs to a linear subspace $\uSpace$ of a Hilbert space $\vSpace$ with norm $\norm{\cdot}{\vSpace}$, dual space $\vSpace'$, and which is continuously embedded into the Lebesgue space $L^2(\Omega;\bR^n)$;
\item\label{ass:pressure} $p(t)$ belongs to a Hilbert space $\pSpace$ continuously embedded into $L^2(\Omega)$.
\end{enumerate} 
For typical flows, $\vSpace$ encodes regularity and boundary conditions, while $\uSpace$ includes further constraints such as incompressibility. In the body-forced flow considered in the introduction, for instance, $\vec{u}$ is simply the velocity field, $\vSpace=H^1_0(\Omega; \bR^n)$ is the Sobolev space of (weakly) differentiable square-integrable vector fields that vanish on the domain boundary and have square-integrable derivatives, $\uSpace$ is its divergence-free subspace, and $\pSpace = L^2(\Omega)$. In the Rayleigh--B\'enard problem discussed in \cref{ex:rb-equations} below, instead, $\vec{u}$ represents both the fluid's velocity and its temperature. Note that assumption~\cref{ass:subspaces} requires all boundary conditions defining $\vSpace$ and all other constraints defining  $\uSpace$ to be homogeneous; this is usually achieved by letting $\vec{u}(t)$ and $p(t)$ be perturbations from a reference flow state (say, a laminar flow). 

The state of the fluid evolves according to a differential equation in the form
\begin{equation}\label{e:flow-eq}
\ddt{\vec{u}} = \opA  \vec{u} + \opB(\vec{u},\vec{u}) + \opC p +  \vec{f}, \qquad \vec{u}(0)=\vec{u}_0,
\end{equation}
where
$\opA:\vSpace \to \vSpace'$ is a linear differential operator that typically represents the effects of diffusion, advection by a mean flow, and buoyancy forces proportional to the temperature of the fluid;
$\opB: \vSpace \times \vSpace \to \vSpace'$ is a bilinear differential operator representing nonlinear advection;
$\opC: \pSpace \to \vSpace'$ is linear differential operator representing pressure forces;
$\vec{f} \in L^2(\Omega; \bR^n)$ is an externally applied steady force\footnote{The methods discussed in this work can be extended to the case of time-dependent external forces if an explicit bound on the spatial $L^2$ norm $\norm{\vec{f}(t)}{2}$ is available pointwise in time, but we do not consider this for brevity.};
and $\vec{u}_0 \in \uSpace$ is the initial state of the fluid. Equation~\cref{e:flow-eq} is essentially an abstract version of the usual momentum equation~\cref{e:intro-ns-momentum}, where $\opA \vec{u} = \Rey^{-1}\Delta \vec{u}$, $\opB(\vec{u},\vec{u})=-\vec{u}\cdot \nabla \vec{u}$ and $\opC p = -\nabla p$. Such an abstraction enables us to consider flows that require evolution equations for quantities beyond the fluid's velocity, such as its temperature or vorticity, whilst using a compact notation.

The operators $\opA$, $\opB$ and $\opC$ are associated to a bounded\footnote{A $k$-linear form $f:\vSpace \times \cdots \times \vSpace \to \bR$ is bounded if $\abs{f(\vec{u}_1,,\ldots,\vec{u}_k)} \leq C \norm{\vec{u}_1}{\vSpace} \cdots \norm{\vec{u}_k}{\vSpace}$ for some constant $C>0$ and all $\vec{u}_1,\ldots,\vec{u}_k\in \vSpace$.} 
bilinear form $a:\vSpace \times \vSpace \to \bR$ (not necessarily symmetric),
a bounded trilinear form $b:\vSpace \times \vSpace \times \vSpace \to \bR$, 
and a bounded bilinear form $c:\vSpace \times \pSpace \to \bR$
such that
\begin{subequations}
\label{e:abstract-forms-abc}
\begin{align}
	\label{e:bilinear-form-a}
	&&&&&&\ip{\vec{u}_2}{\opA \vec{u}_1} &= a( \vec{u}_2, \vec{u}_1) &&\forall  \vec{u}_1, \vec{u}_2 \in \vSpace,  &&&&&&\\
	\label{e:trilinear-form-b}
	&&&&&&\ip{\vec{u}_2}{\opB( \vec{u}_1, \vec{u}_1 ) } &= b( \vec{u}_2, \vec{u}_1, \vec{u}_1) &&\forall  \vec{u}_1 \in \uSpace, \vec{u}_2 \in \vSpace,\\
	\label{e:pressure-form-c}
	&&&&&&\ip{\vec{u}}{\opC p } &= c( \vec{u}, p) &&\forall  \vec{u}\in \vSpace, \, p \in \pSpace,
\end{align}
\end{subequations} 
where angled brackets denote the inner product in $L^2(\Omega;\bR^n)$:
\begin{equation}
\ip{\vec{u}_1}{\vec{u}_2} := \int_\Omega \vec{u}_1 \cdot \vec{u}_2 \, \dVolume \qquad \forall \vec{u}_1, \vec{u}_2 \in L^2(\Omega;\bR^n).
\end{equation}
For the body-forced flow considered in the introduction, for example, $a( \vec{u}_2, \vec{u}_1) = \Rey^{-1}\int_{\Omega} \nabla \vec{u}_1 \cdot \nabla \vec{u}_2 \dVolume$, $b( \vec{u}_2, \vec{u}_1, \vec{u}_1) = \int_{\Omega} \vec{u}_1 \cdot \nabla \vec{u}_2 \cdot \vec{u}_1 \dVolume$ and $c( \vec{u}, p) = \int_{\Omega} (\nabla \cdot \vec{u}) p \, \dVolume$. Recalling that for this example $\vSpace=H^1_0(\Omega; \bR^n)$, $\uSpace$ is the divergence-free subspace of $\vSpace$, and $\pSpace=L^2(\Omega)$, the identities (\ref{e:abstract-forms-abc}a--c) can be verified using integration by parts, the no-slip boundary conditions of functions in $\vSpace$ and, for~\cref{e:trilinear-form-b}, the incompressibility of $\vec{u}_1 \in \uSpace$. 

It is also useful to consider an alternative representation for the trilinear form $b$. For this, we define a linear operator $\opH$ mapping a vector field $\vec{u}_1 \in \vSpace$ into a linear operator $\opH(\vec{u}_1):\vSpace \to \vSpace'$ via the identity 
\begin{equation}\label{e:opH}
\ip{ \vec{u}_2 }{ \opH(\vec{u}_1) \vec{u}_3 } = b( \vec{u}_1, \vec{u}_2, \vec{u}_3)    
\qquad \forall \vec{u}_2, \vec{u}_3 \in \vSpace.
\end{equation}
For the body-forced flow we may take $\opH$ to satisfy $\opH(\vec{u}_1)\vec{u}_3 = \nabla \vec{u}_1 \cdot \vec{u}_3$, and \cref{ex:rb-equations} provides the relevant definition for Rayleigh--B\'enard convection.

Finally, to model the energy conservation properties of the nonlinear advection and pressure terms in Navier--Stokes equations, we make two technical assumptions that are usually satisfied by incompressible flows subject to no-slip, periodic, or stress-free boundary conditions:
\begin{enumerate}[({A}1), resume, leftmargin=*, labelsep=1em, widest=9, topsep=1.5ex, noitemsep=0ex]
\item\label{ass:nnl-energy-conservation} There exists a linear space $\energyPreservingSpace$ of symmetric $n \times n$ matrices such that
$b(\Lambda \vec{u},\vec{u},\vec{u}) = 0$ for all $\Lambda \in \energyPreservingSpace$ and all $\vec{u} \in \uSpace$.
\item\label{ass:pressure-conservation} There exists a linear subspace $\energyPreservingSpace_0 \subseteq \energyPreservingSpace$ such that
$c(\Lambda \vec{u},p) = 0$ for all $\Lambda \in \energyPreservingSpace_0$, $\vec{u} \in \uSpace$ and $p \in \pSpace$.
\end{enumerate}
In the case of body-forced flows governed by~\cref{e:intro-ns}, for example, $\energyPreservingSpace \equiv \energyPreservingSpace_0$ contains all multiples of the identity matrix, since $b(\lambda \vec{u},\vec{u},\vec{u}) = \lambda \int_{\Omega} \vec{u} \cdot \nabla \vec{u} \cdot \vec{u} \dVolume=0$ and $c(\lambda \vec{u},p) = \lambda \int_{\Omega} (\nabla \cdot \vec{u}) p \dVolume = 0$ by virtue of incompressibility and the no-slip boundary conditions. For the Rayleigh--B\'enard problem discussed in \cref{ex:rb-equations} below, instead, the spaces $\energyPreservingSpace$ and $\energyPreservingSpace_0$ are more general and do not coincide.

Within this abstract framework, we seek \textit{a priori} upper and lower bounds on the infinite-time average of a quantity 
that can be represented by a bounded bilinear form $\Phi:\vSpace \times \vSpace \to \bR$ (not necessarily symmetric) and is associated to a linear operator $\opP:\vSpace \to \vSpace'$ via
\begin{equation}\label{e:general-phi-def}
\Phi(\vec{u}, \vec{u}) := \ip{\vec{u}}{\opP \vec{u}}.
\end{equation}
One example is the energy dissipation for the body-forced flow considered in the introduction, which is given by $\Phi(\vec{u},\vec{u}) = \int_{\Omega} \abs{\nabla \vec{u}}^2 \dVolume$ and can be expressed in the inner-product form~\cref{{e:general-phi-def}} with $\opP \vec{u} = - \Delta \vec{u}$ upon integration by parts. 
Since the infinite-time average of $\Phi(\vec{u}(t),\vec{u}(t))$ may depend on the initial state $\vec{u}(0)$, we focus on bounding from above the maximal average
\begin{equation}
\maxAverage := \sup_{\vec{u}(0) \in \uSpace} \overline{\Phi(\vec{u}(t), \vec{u}(t))},
\end{equation}
where overbars denote averaging over infinite time:
\begin{equation}
\overline{\Phi(\vec{u}(t), \vec{u}(t))} := \limsup_{T \to \infty} \frac{1}{T}\int_0^T \Phi(\vec{u}(t),\vec{u}(t)) \, {\rm d}t.
\end{equation}
Lower bounds on the minimal average are of equal interest, but can be deduced from upper bounds on the maximal average of $-\Phi$ and will therefore not be discussed.

\begin{example}\label{ex:rb-equations}
Consider the problem of bounding the mean vertical heat flux in Rayleigh--B\'enard (RB) convection between stress-free plates. Using standard nondimensional variables (see, e.g.,~\cite{Goluskin2016book}) and writing $\vec{x}=(x,y,z)$ for the position vector, the fluid occupies the domain $\Omega = [-L_x,L_x] \times [-L_y,L_y] \times [0,1]$, which we assume to be periodic in the horizontal directions ($x$ and $y$). The bottom plate ($z=0$) is held at a constant temperature $T=1$, while the top one ($z=1$) is held at $T=0$. 

To lift the inhomogeneous boundary conditions on the temperature, we consider the state vector $\vec{u}(t)=(\vec{w}(t),\theta(t))$ that describes incompressible velocity perturbations $\vec{w}=(w_1,w_2,w_3)$ and temperature perturbations $\theta$ from a purely conductive state, with no fluid motion and temperature distribution $1-z$. The flow is then described by the nondimensional Boussinesq equations in perturbation form,
\begin{equation}\label{e:rb-equation-example}
	\ddt{}
	\begin{pmatrix}\vec{w}\\ \theta\end{pmatrix}
	=
	\underbrace{
		\begin{pmatrix} \Pran \, \Delta\vec{w} + \Pran \Ra \, \theta \,\vec{e}_3 \\ \Delta \theta + \vec{w} \cdot\vec{e}_3 \end{pmatrix}
	}_{\opA \vec{u}}
	+
	\underbrace{
		\begin{pmatrix} -\vec{w} \cdot \nabla \vec{w} \\ -\vec{w} \cdot \nabla \theta \end{pmatrix}
	}_{\opB(\vec{u},\vec{u})}
	+
	\underbrace{
		\begin{pmatrix} -\nabla p \\ 0 \end{pmatrix}
	}_{\opC p},
\end{equation}
where the Prandtl number $\Pr$ is the ratio of kinematic viscosity and thermal diffusivity, the Rayleigh number $\Ra$ describes the strength of the thermal forcing, $\vec{e}_3$ is the unit vector in the vertical direction, and there is no external forcing ($\vec{f}=0$). 
The instantaneous vertical heat flux is
\begin{equation}\label{e:rb-heat-flux}
	\Phi(\vec{u}(t),\vec{u}(t)) 
	= \int_\Omega \theta(t,\vec{x}) \, w_3(t,\vec{x}) \, \dVolume 
	= \int_\Omega \vec{u}(t,\vec{x}) \cdot \underbrace{\begin{pmatrix} 0 & 0 \\ \vec{e}_3^\tr & 0	\end{pmatrix} \vec{u}(t,\vec{x})}_{\opP \vec{u}} \, \dVolume.
\end{equation}

Suitable pressure and velocity spaces are 
$\pSpace = L^2(\Omega)$, 
$\vSpace = \wSpace \times H^1_{p,0}(\Omega)$ with $\wSpace = H^1_{p}(\Omega)\times H^1_{p}(\Omega) \times H^1_{p,0}(\Omega)$, and
$\uSpace = \{ (\vec{w},\theta) \in \vSpace : \nabla \cdot \vec{w} = 0 \}$, 
where $H^1_{p}(\Omega)$ is the space of horizontally-periodic square-integrable functions with square-integrable weak derivatives and $H^1_{p,0}$ is its subspace of functions that vanish at $z=0$ and $z=1$.
Writing $\vec{v}=(\vecphi,\tau)$ with $\vecphi = (\phi_1,\phi_2,\phi_3)$, the forms $a$, $b$ and $c$ in~\cref{e:abstract-forms-abc} are
\begin{subequations}
	\begin{gather}
		\label{e:rb-example-a}
		a(\vec{v},\vec{u}) = 
		\int_{\Omega} \left(
		-\Pr\, \nabla \vecphi \cdot \nabla \vec{w}
		+\Pr\Ra\,\theta \phi_3
		+ \tau w_3
		- \nabla \theta \cdot \nabla \tau 
		\right) \dVolume,\\
		\label{e:rb-example-b}
		b(\vec{v},\vec{u},\vec{u}) = 
		\int_{\Omega} \left(
		\vec{w} \cdot \nabla \vecphi \cdot \vec{w} + \nabla\tau \cdot \vec{w} \theta
		\right) \dVolume,\\
		\label{e:rb-example-c}
		c(\vec{v},p) = 
		\int_{\Omega} (\nabla \cdot \vecphi) \,p \,\dVolume.
	\end{gather}
\end{subequations}
These expression are obtained by taking the inner product of $\vec{v} \in \vSpace$ with the right-hand side of~\cref{e:rb-equation-example}, integrating some terms by parts using the divergence-free and boundary conditions under the assumption that $(\vec{w},\theta) \in \uSpace$, and recognizing that the result is well defined for all $(\vec{w},\theta) \in \vSpace$. 
From~\cref{e:rb-example-b}, one sees that the action of the operator $\opH$ defined via~\cref{e:opH} is $\opH(\vec{v})\vec{u} = \left( \nabla \vecphi \cdot \vec{w},\, \nabla \tau \cdot \vec{w}\right)$.
Finally, assumptions~\cref{ass:nnl-energy-conservation} and~\cref{ass:pressure-conservation} hold with
\begin{equation}\label{e:rb-balance-parameter-sets}
	\energyPreservingSpace = \left\{ \Lambda =  \begin{pmatrix}
		\frac{\lambda_1}{\Pran \Ra} I & \lambda_3 \vec{e}_3\\ \lambda_3 \vec{e}_3^\tr & \lambda_2 \end{pmatrix}: 
	\lambda_i \in \bR \right\}
	\quad \text{and} \quad
	\energyPreservingSpace_0 = \{\Lambda \in \energyPreservingSpace: \,\lambda_3 = 0 \},
\end{equation}
where we divide $\lambda_1$ by $\Pran \Ra$ to ease the notation in subsequent examples. Indeed, integration by parts using the divergence-free and the boundary conditions imply that $\int_{\Omega} \vec{w} \cdot \nabla\vec{w} \cdot \vec{w} \,\dVolume = 0$, $\int \nabla \theta \cdot \vec{w} \theta \,\dVolume = 0$, and $\int_{\Omega} \nabla w_3 \cdot \vec{w}\theta \,\dVolume = - \int_{\Omega} \vec{w} \cdot \nabla(\theta \vec{e}_3) \cdot \vec{w} \, \dVolume$. Thus, for any $\Lambda$ in the space $\energyPreservingSpace$ specified above and any $(\vec{w},\theta) \in \uSpace$, the quantity
\begin{equation}
	b(\Lambda \vec{u},\vec{u},\vec{u}) 
	= \int_{\Omega} \left(
	\vec{w} \cdot \nabla \left(\tfrac{\lambda_1 \vec{w}}{\Pran \Ra} + \lambda_3 \theta \vec{e}_3\right) \cdot \vec{w} + \nabla(\lambda_3 w_3 + \lambda_2 \theta) \cdot \vec{w} \theta
	\right) \dVolume
\end{equation}
vanishes identically as required. Similarly, since $\vec{w}$ is divergence-free, the quantity
\begin{equation}
	c(\Lambda \vec{u},p) 
	= \int_{\Omega} \nabla \cdot \left(\tfrac{\lambda_1 \vec{w}}{\Pran \Ra} + \lambda_3 \theta \vec{e}_3\right) \,p \,\dVolume
	= \lambda_3 \int_{\Omega}  p\, \partial_z \theta \,\dVolume
\end{equation}
vanishes identically for all $(\vec{w},\theta) \in \uSpace$ and all $p \in \pSpace$ only if $\lambda_3=0$, so $\Lambda \in \energyPreservingSpace_0$.
\end{example}

\section{The background method via auxiliary functions}\label{ss:general-background-method}

Upper bounds on the maximal infinite-time average $\maxAverage$ can be derived using a general strategy based on the following simple observation: the time derivative of any bounded and differentiable function $V:\uSpace \to \bR$, which we refer to as an \textit{auxiliary function}, has zero infinite-time average. 
Specifically, let $\LieDer V:\uSpace \times \pSpace \to \bR$ be the Lie derivative of $V$ along solutions of~\cref{e:flow-eq}, meaning that $\LieDer V(\vec{u}(t), p(t)) = \ddt{}V(\vec{u}(t))$. By the fundamental theorem of calculus, the infinite-time average of $\LieDer V(\vec{u}(t), p(t))$ vanishes if $V(\vec{u}(t))$ is bounded uniformly in time, so 
\begin{equation}\label{e:av-identity}
\overline{\Phi(\vec{u}(t),\vec{u}(t))} = \overline{\Phi(\vec{u}(t),\vec{u}(t)) + \LieDer V(\vec{u}(t), p(t))}
\end{equation}
irrespective of the initial state $\vec{u}(0)$. If $V$ can be chosen such that
\begin{equation}\label{e:af-general-constraint}
\Phi(\vec{u},\vec{u}) + \LieDer V(\vec{u}, p) \leq \gamma \qquad \forall (\vec{u}, p) \in \uSpace \times \pSpace
\end{equation}
for some constant $\gamma$, then the right-hand side of~\cref{e:av-identity} is also bounded above by $\gamma$ irrespective of the initial state, and we conclude that $\maxAverage \leq \gamma$.

The background method amounts to searching for the smallest possible bound $\gamma$ over \textit{quadratic} auxiliary functions $V$ constructed to ensure that inequality \cref{e:af-general-constraint} is quadratic in $\vec{u}$ and $p$~\cite{Chernyshenko2014a,Chernyshenko2017}. The body-forced flow example given in the Introduction, for instance, uses the auxiliary function $V(\vec{u})=\frac{\lambda}{2}\int_{\Omega} \abs{\vec{u}-\vec{U}}\dVolume$, which represents the energy of perturbations from an incompressible background field $\vec{U}$, scaled by the balance parameter $\lambda$. In this case, the general inequality~\cref{e:af-general-constraint} reduces exactly to~\cref{e:intro-bounding-ineq} and the pressure drops out by virtue of the incompressibility of $\vec{U}$. \Cref{ss:basic-formulation} extends this example, showing how generalized background fields and generalized perturbation energies can be used to derive bounds for any flow whose governing equations fit within the abstract setting introduced in \cref{s:governing-equations}. \Cref{ss:symmetry-reduction}, instead, discusses how symmetries in the governing equations and in $\Phi$ can be exploited when optimizing~$\gamma$. Finally, \cref{ss:additional-constraints} outlines how additional constraints on the asymptotic behaviour of the flow can be taken into account to try and improve the optimal bound.

\subsection{General formulation}\label{ss:basic-formulation}

The background method considers quadratic auxiliary functions of the form
\begin{equation}\label{e:bm-af}
V(\vec{u}) = \tfrac12 \ip{\Lambda \vec{u}}{\vec{u}} - \ip{\vec{v}}{\vec{u}},
\end{equation}
where $\vec{v} \in \vSpace$ is a time-independent vector field and $\Lambda \in \energyPreservingSpace$ is a symmetric matrix satisfying assumption~\cref{ass:nnl-energy-conservation}. 
Traditional formulations of the method~\cite{Doering1994,Constantin1995,Doering1996}, including that given in the Introduction for body-forced flows, additionally take $\vec{v} = \Lambda \backgroundField$ for some $\backgroundField \in \uSpace$. In this case, one may rewrite the $V$ in~\cref{e:bm-af} as
\begin{equation}
V(\vec{u}) 
= \tfrac12 \ip{\Lambda (\vec{u}-\backgroundField)}{\vec{u}-\backgroundField}
+ \tfrac12 \ip{\Lambda \backgroundField}{\backgroundField}
\end{equation}
and interpret this as the ``energy'' of perturbations from the \textit{background field} $\backgroundField$, weighted by the \textit{balance parameter} $\Lambda$, plus a constant term that could be dropped if desired because it gives no contribution to the Lie derivative $\mathcal{L}V$ appearing in~\cref{e:af-general-constraint}.
Here, however, we remove the restriction that $\vec{v}=\Lambda\backgroundField$ and consider arbitrary $\vec{v} \in \vSpace$, which will be called the \textit{generalized background field}. This generalization is convenient for two reasons. First, it may improve the best upper bound on $\maxAverage$ that can be proven, because the vectors $\vec{v}=\Lambda\backgroundField$ generated by classical background fields usually span only a strict subspace of $\vSpace$. Second, as shown below, minimizing the upper bound $\gamma$ on $\maxAverage$ over $\Lambda \in \energyPreservingSpace$ and $\vec{v} \in \vSpace$ is a convex problem, while doing so over $\Lambda \in \energyPreservingSpace$ and $\backgroundField \in \uSpace$ is not. This convexity allows for the efficient computational optimization of $\gamma$ using the approaches described in \cref{s:implementation}.

Differentiating~\cref{e:bm-af} along solutions of~\cref{e:flow-eq}, using the identities~(\ref{e:abstract-forms-abc}a--c), and invoking assumption~\cref{ass:nnl-energy-conservation} yields
\begin{align}
\Phi(\vec{u},\vec{u}) + \LieDer V(\vec{u}, p) 
&= \Phi(\vec{u},\vec{u}) + \ip{ \Lambda \vec{u}  - \vec{v} }{\,\opA \vec{u} + \opB(\vec{u},\vec{u}) + \opC p + \vec{f} }
\nonumber\\
&= c(\Lambda \vec{u} - \vec{v}, p) - \ip{\vec{v}}{\vec{f}} - 2\linear(\vec{u}) - \quadratic(\vec{u},\vec{u}),
\end{align}
where $\quadratic$ and $\linear$ are quadratic and linear forms, respectively, defined as
\begin{subequations}
\begin{align}
	\label{e:homogeneous-quadratic-term}
	\quadratic(\vec{u},\vec{u}) &:= b(\vec{v},\vec{u},\vec{u}) - a(\Lambda \vec{u}, \vec{u}) - \Phi(\vec{u},\vec{u}),\\
	\linear(\vec{u}) &:= \tfrac12 a(\vec{v},\vec{u}) - \tfrac12 \ip{\vec{f}}{\Lambda\vec{u}}.
	\label{e:linear-term}
\end{align}
\end{subequations}
Inequality~\cref{e:af-general-constraint} then reduces to
\begin{equation}\label{e:af-quadratic-constraint}
c(\Lambda \vec{u} - \vec{v}, p) 
- \ip{\vec{v}}{\vec{f}}  
- 2\linear(\vec{u}) 
- \quadratic(\vec{u},\vec{u}) 
\leq  \gamma  \quad \forall (\vec{u}, p) \in \uSpace \times \pSpace,
\end{equation}
and the smallest value of $\gamma$ for which this condition holds is the best upper bound on $\maxAverage$  that can be proven with the background method as formulated in this section. As anticipated above, minimizing $\gamma$ over the generalized background field $\vec{v}$ and balance parameter $\Lambda$ is a convex problem because~\cref{e:af-quadratic-constraint} is linear in the optimization variables. 

For fixed $\Lambda$ and $\vec{v}$, the best choice of $\gamma$ clearly coincides with the supremum of the left-hand side of~\cref{e:af-quadratic-constraint} over $\uSpace \times \pSpace$. Since the dependence on $p$ is linear, this supremum is finite only if  $c(\Lambda \vec{u} - \vec{v}, p)$ vanishes for all $\vec{u} \in \uSpace$ and $p \in \pSpace$. In particular, we must have $c(\vec{v},p)=\ip{\vec{v}}{\opC p} = \ip{\opC^\star \vec{v}}{p} = 0$ for all $p \in \pSpace$, where $\opC^\star$ is the adjoint of $\opC$, and $c(\Lambda \vec{u}, p)=0$ for all $p \in \pSpace$ and all $\vec{u} \in \uSpace$. The former condition requires $\opC^\star \vec{v} = 0$, which in typical applications constrains the velocity component of $\vec{v}$ to be incompressible (see \cref{ex:rb-bm-general} below for details in the context of RB convection). The latter, instead, can be ensured by taking $\Lambda \in \energyPreservingSpace_0$ according to assumption~\cref{ass:pressure-conservation}.  These observations can be summarized as follows.

\begin{theorem}[General background method]\label{th:optimal-bm-general}
The best upper bound $\gamma^*$ on $\maxAverage$ provable with the background method is
\begin{subequations}
	\begin{align}\label{e:bound-sdp-form}
		\gamma^* 
		&= \mathop{\vphantom{p}\inf}\limits_{\substack{(\Lambda, \vec{v}) \in \energyPreservingSpace_0  \times \vSpace \\ \opC^\star \vec{v}= 0}}\;
		\left\{\gamma : \; 
		- \ip{\vec{v}}{\vec{f}} - 2\linear(\vec{u}) - \quadratic(\vec{u},\vec{u})
		\leq  \gamma  \;\; \forall \vec{u} \in \uSpace \right\}
		\\
		&=\mathop{\vphantom{p}\inf}\limits_{\substack{(\Lambda, \vec{v}) \in \energyPreservingSpace_0  \times \vSpace \\ \opC^\star \vec{v}= 0}} \; 
		\sup_{\substack{\vec{u} \in \uSpace}} 
		\;\left\{ - \ip{\vec{v}}{\vec{f}} - 2\linear(\vec{u}) - \quadratic(\vec{u},\vec{u}) \right\}.
		\label{e:bound-saddle-point-form}
	\end{align}
\end{subequations}
\end{theorem}

Next, we state necessary and sufficient condition for this upper bound to be finite.

\begin{proposition}[Spectral constraint]\label{prop:spectral-constraint}
The upper bound on $\maxAverage$ in \cref{th:optimal-bm-general} is finite if and only if there exist $\Lambda \in \energyPreservingSpace_0$ and $\vec{v} \in \vSpace$ satisfying $\opC^\star \vec{v}=0$ and such that:
\begin{enumerate}[(i), leftmargin=*, labelsep=1em, widest=ii, noitemsep, topsep=1ex]
	\item\label{sc:compatibility} $\ell(\vec{u}) = 0$ whenever $q(\vec{u},\vec{u})=0$;
	\item\label{sc:nonnegativity} $q(\vec{u},\vec{u})$ satisfies
	\begin{equation}\label{e:spectral-constraint}
		\inf_{\substack{\vec{u} \in \uSpace:\; \left\|\vec{u}\right\|_2 = 1}} \quadratic(\vec{u},\vec{u}) \geq 0.
	\end{equation}
\end{enumerate}
\end{proposition}

\begin{remark}\label{remark:spectral-constraint}
Condition~\cref{sc:nonnegativity} in \cref{prop:spectral-constraint} is a \textit{spectral constraint}, so called because it requires the nonnegativity of all (real) eigenvalues $\mu$ of the self-adjoint linear eigenvalue problem
\begin{equation}\label{e:general-eig-prob-sc}
	\left( \opH(\vec{v}) + \opH(\vec{v})^\star - \Lambda \opA - \opA^\star \Lambda - \opP - \opP^\star \right)\vec{u} = \mu \vec{u},
\end{equation}
where the linear operators appearing on the left-hand side are defined in \cref{s:governing-equations} and stars indicate adjoints (see, e.g.,~\cite{Doering1994,Constantin1995,Doering1996}). The corresponding eigenfunctions are subject to all constraints embedded in the definition of the space $\uSpace$, plus any additional ``natural'' boundary conditions that arise when deriving~\cref{e:general-eig-prob-sc} as a necessary optimality condition for $\vec{u}$ in~\cref{e:spectral-constraint}. Condition~\cref{sc:compatibility} in \cref{prop:spectral-constraint}, instead, is a solvability condition for the Euler--Lagrange equation characterizing the optimal $\vec{u}$ for problem~\cref{e:bound-saddle-point-form},
\begin{equation}
	\left( \opH(\vec{v}) + \opH(\vec{v})^\star - \Lambda \opA - \opA^\star \Lambda - \opP - \opP^\star \right)\vec{u} 
	= \Lambda \vec{f} - \opA^\star \vec{v}.
\end{equation}
\end{remark}

\begin{proof}[Proof of \cref{prop:spectral-constraint}]
To see that conditions~\cref{sc:compatibility,,sc:nonnegativity} suffice to obtain a finite bound on $\maxAverage$, observe that either all $\vec{u}$-dependent terms in~\cref{e:bound-saddle-point-form} vanish, or the nonnegative quadratic term dominates as $\|\vec{u}\|_2$ increases. For the necessity of~\cref{sc:nonnegativity}, note that the optimal bound on $\maxAverage$ stated in \cref{th:optimal-bm-general} cannot be finite unless $\quadratic(\vec{u},\vec{u})$ is bounded below. Recalling from~\cref{e:homogeneous-quadratic-term,e:general-phi-def,e:bilinear-form-a,e:trilinear-form-b} that $\quadratic$ is a homogeneous bilinear form, the only possible finite lower bound is zero, and it suffices to establish it for $\vec{u}$ with unit $L^2$ norm. Condition~\cref{sc:compatibility}, instead, is necessary because if $\quadratic(\vec{u}_0,\vec{u}_0)=0$ but $\ell(\vec{u}_0)\neq 0$, then one may set $\vec{u} = k \vec{u}_0$ and select the constant $k$ to make $\ip{\vec{v}}{\vec{f}} - 2\linear(\vec{u}) - \quadratic(\vec{u},\vec{u})= \ip{\vec{v}}{\vec{f}} - 2k \linear(\vec{u}_0)$ arbitrarily large in~\cref{e:bound-saddle-point-form}. 
\end{proof}

\begin{example}\label{ex:rb-bm-general}
Let us apply the approach described above to bound the mean vertical heat transport in the RB convection problem introduced in \cref{ex:rb-equations}.
With a balance parameter $\Lambda$ in the admissible set $\energyPreservingSpace$ in~\cref{e:rb-balance-parameter-sets} and a generalized background field $\vec{v}=(\vecphi,\tau) \in \vSpace$, the quadratic auxiliary function in~\cref{e:bm-af} becomes
\begin{equation}
	V(\vec{u}) 
	= \frac12
	\underbrace{\int_{\Omega} \left( \frac{\lambda_1}{\Pran \Ra} \abs{\vec{w}}^2 + \lambda_2 \abs{\theta}^2 + 2\lambda_3 \theta w_3 \right) \dVolume}_{\ip{\Lambda \vec{u}}{\vec{u}}}
	- \underbrace{\int_{\Omega} \left( \vecphi \cdot \vec{w} + \tau \theta \right)\dVolume}_{\ip{\vec{v}}{\vec{u}}}.
\end{equation}
This can be rewritten as
\begin{align}
	V(\vec{u}) = 
	&\;\frac{\lambda_1}{2\Pran\Ra} \int_{\Omega} \Big\vert \vec{w} - \frac{\Pran\Ra}{\lambda_1} \vecphi \Big\vert^2 \dVolume 
	+ \frac{\lambda_2}{2} \int_{\Omega} \Big\vert \theta - \frac{1}{\lambda_2} \tau \Big\vert^2 \dVolume
	\nonumber\\
	&+ \lambda_3 \int_\Omega w_3 \theta \,\dVolume
	- \frac{(\Pran\Ra)^2}{2\lambda_1} \int_{\Omega} \abs{\vecphi}^2 \dVolume
	- \frac{1}{2\lambda_2} \int_\Omega \tau^2 \,\dVolume,
	\label{e:rb-example-af}
\end{align}
where the first three terms represent the energy of deviations from the background velocity $(\Pran\Ra/\lambda_1) \vecphi$, the energy of perturbations from the background temperature $\tau/\lambda_2$, and the vertical convective heat flux.
Using the definition of the instantaneous vertical heat transport $\Phi$ in~\cref{e:rb-heat-flux} and the forms $a$, $b$ and $c$ in~\cref{e:rb-example-a}--\cref{e:rb-example-c}, inequality~\cref{e:af-quadratic-constraint} becomes
\begin{align}\label{e:rb-example-constraint}
	\int_{\Omega}\left(
	\lambda_3 \abs{w_3}^2 
	+\lambda_3 \Pran\Ra \abs{\theta}^2
	-\tfrac{\lambda_1}{\Ra} \abs{\nabla \vec{w}}^2 
	-\lambda_2 \abs{\nabla \theta}^2 
	- \vec{w}\cdot \nabla\vecphi \cdot \vec{w} 
	\right) \,&\dVolume
	\nonumber\\
	+ \int_{\Omega} \left( [(1 + \lambda_1 + \lambda_2)\vec{e}_3 - \nabla\tau]\cdot \vec{w}\theta
	- \lambda_3	(\Pran+1)\nabla\theta \cdot \nabla w_3
	\right) \,&\dVolume
	\nonumber\\
	+ \int_{\Omega} \left(
	\Pr \nabla\vecphi \cdot \nabla\vec{w} - \Pr\Ra \theta \phi_3 - \tau w_3 + \nabla\tau \cdot \nabla\theta
	\right) \,&\dVolume
	\nonumber\\
	+\int_{\Omega}\left( \lambda_3 \partial_z \theta - \nabla \cdot \vecphi \right) p  \;&\dVolume
	\;\leq\;\gamma.
\end{align}
This must hold, by a suitable choice of $\gamma$, $\lambda_1$, $\lambda_2$, $\lambda_3$, $\tau$ and $\vecphi$, for all velocity and temperature perturbations $(\vec{w},\theta) \in \uSpace$ and all pressure fields $p \in \pSpace$.

Forcing the pressure terms to vanish yields $\opC^\star \vec{v} := \nabla \cdot \vecphi = 0$, meaning that the velocity component of the generalized background field must be incompressible, and $\lambda_3=0$, so the balance parameter $\Lambda$ belongs to the space $\energyPreservingSpace_0$ in~\cref{e:rb-balance-parameter-sets}.	With these simplifications, the spectral constraint~\cref{e:spectral-constraint} requires
\begin{equation}\label{e:rb-sc-integral}
	\int_{\Omega}\!\left(\!
	\tfrac{\lambda_1}{\Ra} \abs{\nabla \vec{w}}^2 + \lambda_2 \abs{\nabla \theta}^2
	+ [\nabla\tau - (1+\lambda_1+\lambda_2)\vec{e}_3]\cdot \vec{w}\theta
	+ \vec{w}\cdot \nabla\vecphi \cdot \vec{w}\!
	\right)\!\dVolume \geq 0
\end{equation}
for all $(\vec{w},\theta) \in \uSpace$ with unit $L^2$ norm, i.e., $\int_\Omega (\abs{\vec{w}}^2 + \abs{\theta}^2 )\, \dVolume = 1$. This is true if and only if the eigenvalue problem
\begin{subequations}
	\begin{gather}
		-2\lambda_2\, \Delta\theta + [\nabla\tau - (1+\lambda_1+\lambda_2)\vec{e}_3] \cdot \vec{w} = \mu \theta\\
		- \frac{2\lambda_1}{\Ra} \Delta\vec{w} + (\nabla\vecphi + \nabla\vecphi^\tr) \vec{w} + [\nabla\tau - (1+\lambda_1+\lambda_2)\vec{e}_3] \theta + \nabla p = \mu \vec{w}\\
		\nabla \cdot \vec{w} = 0
	\end{gather}
\end{subequations}
has only nonnegative eigenvalues $\mu$, where the ``pressure'' $p$ is a Lagrange multiplier that enforces the incompressibility of $\vec{w}$. 
\end{example}

\begin{remark}
Requiring $\lambda_3=0$ in \cref{ex:rb-bm-general} means that including the flux $\int_{\Omega} w_3 \theta \, \dVolume$ in~\cref{e:rb-example-af} does not help when trying to bound the mean vertical convective heat flux in RB convection with the approach described in this section. Similar flux terms, however, are necessary in double-diffusive convection~\cite{Balmforth2006}. Moreover, they can be exploited even in RB convection by enforcing additional constraints on the admissible flow states~\cite{Tilgner2019}. This is discussed in \cref{ss:additional-constraints}.
\end{remark}

\subsection{Symmetry reduction}\label{ss:symmetry-reduction}

Analysis and numerical treatment of the optimization problem~\cref{e:bound-sdp-form} for the best bound available to the background method can be considerably simplified in the presence of symmetries. Such symmetries may include, but are not limited to, reflection of the flow variables about a symmetry plane and translation invariance for flows on periodic domains. Abstractly, a symmetry can be described using a group $\group$ of orthogonal linear transformations $(\vec{u},p) \mapsto \GroupAction(\vec{u},p) = (\GroupAction_1 \vec{u}, \GroupAction_2 p)$ acting on $L^2(\Omega;\bR^{n}) \times \pSpace$. We assume that the subspaces $\uSpace$ and $\vSpace$ are closed under $\group$, meaning that $\vec{u} \in \uSpace$ (resp. $\vSpace$) implies $\GroupAction_1 \vec{u} \in \uSpace$ (resp. $\vSpace$). If the governing equation~\cref{e:flow-eq} and the quantity of interest $\Phi(\vec{u},\vec{u})$ are invariant under $\group$ in a sense made precise by \cref{th:symmetry-reduction} below, then a symmetrization argument similar to that in~\cite[Appendix~A]{Goluskin2019} shows that auxiliary functions used to bound $\maxAverage$ may be taken to be invariant without loss of generality.

\begin{proposition}\label{th:symmetry-reduction}
Let $\group$ be a group of orthogonal linear transformations $(\vec{u},p) \mapsto  \GroupAction(\vec{u},p) = (\GroupAction_1 \vec{u}, \GroupAction_2 p)$. Suppose that:
\begin{enumerate}[{1)}, leftmargin=*, labelsep=1ex, widest=9, topsep=1ex, itemsep=0ex]
	\item $\Phi$ is invariant under $\group$, meaning that $\Phi(\GroupAction_1 \vec{u}, \GroupAction_1 \vec{u}) = \Phi(\vec{u},\vec{u})$ for all $\GroupAction \in \group$.
	\item The operators $\opA$, $\opB$ and $\opC$ in~\cref{e:flow-eq} are equivariant under $\group$, meaning that
	$\opA \GroupAction_1 \vec{u} = \GroupAction_1 \opA\vec{u}$,
	$\opB(\GroupAction_1 \vec{u}, \GroupAction_1 \vec{u} ) = \GroupAction_1 \opB(\vec{u}, \vec{u})$,
	$\opC \GroupAction_2 p = \GroupAction_1 \opC p$ for all $\GroupAction \in \group$.
	\item The forcing $\vec{f}$ is invariant under $\group$, meaning that $\GroupAction_1\vec{f} = \vec{f}$ for all $\GroupAction \in \group$.
\end{enumerate}
Then, any upper bound $\maxAverage \leq \gamma$ provable using an auxiliary function satisfying~\cref{e:af-general-constraint} can also be proven using an auxiliary function that is invariant under $\group$.
\end{proposition}

This result is true for general auxiliary functions, not just the quadratic one in~\cref{e:bm-af} used in the background method. When applied to the latter, it imposes structure to the generalized background field $\vec{v}$ that reduces the number of optimization variables in~\cref{e:bound-sdp-form} and, more importantly, reveals which dynamical information the background method extracts from the governing equation~\cref{e:flow-eq}. The next example, for instance, proves that the generalized background velocity field $\vecphi$ for the RB convection problem in \cref{ex:rb-bm-general} may be set to zero without worsening the bound on the heat transport.\footnote{This fact was known to Charlie Doering, who described it to one of the authors as due to a ``convexity argument''. Unfortunately, we could not find a published reference to this argument.} This confirms rigorously the recent observation~\cite{Ding2020} that the background method as described above cannot exploit the flow's momentum equation beyond the ``energy'' balances encoded by the quadratic term in the auxiliary function~\cref{e:bm-af}.

\begin{example}\label{ex:rb-symmetry}
The Boussinesq equations~\cref{e:rb-equation-example} and the associated spaces $\uSpace$, $\vSpace$ and $\pSpace$ defined in \cref{ex:rb-equations} are invariant under horizontal translations and under the ``flow reversal'' operation
\begin{equation}\label{e:rb-flow-reversal}
	\begin{pmatrix}\vec{w}(\vec{x},t)\\\theta(\vec{x},t)\\p(\vec{x},t)\end{pmatrix}
	\mapsto
	\begin{pmatrix}G\vec{w}(G\vec{x},t)\\\theta(G\vec{x},t)\\p(G\vec{x},t)\end{pmatrix},
	\qquad
	G = \begin{pmatrix}-1 & 0 & 0\\ 0 & -1 & 0\\ 0 & 0 & 1\end{pmatrix}.
\end{equation}
Requiring the quadratic auxiliary function in~\cref{e:rb-example-af} to be invariant under horizontal translation forces the generalized background velocity and temperature fields $\vecphi$ and $\tau$ to depend only on the vertical coordinate $z$. Since $\vecphi$ must be incompressible (cf. \cref{ex:rb-bm-general}) and vanish at the top and bottom boundaries, we must therefore have $\vecphi = (\phi_1(z), \phi_2(z), 0)$. Invariance under the transformation in~\eqref{e:rb-flow-reversal} requires $\vecphi(\vec{x}) = G\vecphi(G\vec{x})$, i.e., $\phi_1(z)=-\phi_1(z)$ and $\phi_2(z)=-\phi_2(z)$. This can be true only if $\vecphi = 0$, proving that the generalized background velocity field may be taken to vanish identically.
\end{example}

\subsection{Imposing additional constraints}\label{ss:additional-constraints}

The bound on $\maxAverage$ obtained in \cref{th:optimal-bm-general} can sometimes be improved by considering further information about the asymptotic behaviour of the flow state $\vec{u}$ and pressure $p$. Indeed, suppose there exists an absorbing subset $\absSet \subsetneq \uSpace \times \pSpace$ in which all solutions $\vec{u}(t), p(t)$ of~\cref{e:flow-eq} remain after an initial transient, or which is at least approached exponentially quickly. Then, $\maxAverage$ is determined by trajectories that start and remain in $\absSet$ for all times, and to derive an upper bound $\gamma$ it suffices to replace constraint~\cref{e:af-quadratic-constraint} with
\begin{equation}\label{e:restricted-bound}
c(\Lambda \vec{u} - \vec{v}, p) - \ip{\vec{v}}{\vec{f}} - 2\linear(\vec{u}) - \quadratic(\vec{u},\vec{u}) \leq  \gamma  
\qquad \forall (\vec{u}, p) \in \absSet.
\end{equation}
Clearly, imposing the inequality on $\absSet$ rather than on the full space $\uSpace \times \pSpace$ cannot worsen the optimal bound $\maxAverage \leq \gamma$ available to the background method.

To ensure that the bound remains rigorous, the set $\absSet$ must be defined using constraints that are derived from the governing equations~\cref{e:flow-eq}. When studying convective flows, for example, one can often invoke extremum principles to enforce uniform pointwise bounds for the fluid's temperature, and one can sometimes consider integral estimates involving the pressure; see~\cite{Arslan2021,Tilgner2019} and \cref{ex:rb-additional-constraints} below for more details. If no rigorous constraints are available, one may also consider sets $\absSet$ defined by reasonable but unproven conditions, such as bounds on the energy spectrum of the flow variables~\cite{Constantin1996a}. In this case, one obtains \textit{conditional} bounds that apply only to solutions of~\cref{e:flow-eq} satisfying the imposed conditions, which may or may not exist.

Irrespective of whether the absorbing set $\absSet$ is defined rigorously or not, we now assume for concreteness that it can be described using one pointwise linear inequality (modelling, say, an extremum principle) and one inequality involving quadratic forms (modelling, say, an integral estimate). Specifically, we let
\begin{align}
\absSet := \big\{
(\vec{u}, p) \in \uSpace \times \pSpace:\quad
&h + \opK \vec{u} + \opL p \geq 0 \text { a.e. on }\Omega,\nonumber \\
&d_0 + d_1(\vec{u},\vec{u}) + d_2(\vec{u},p) + d_3(p,p) \geq 0
\big\},
\label{e:absSet}
\end{align}
where the function $h \in L^2(\Omega)$, the linear operators $\opK: \uSpace \to L^2(\Omega)$ and $\opL: \pSpace \to L^2(\Omega)$, the constant $d_0 \in \bR$, and the bilinear forms $d_1:\uSpace\times\uSpace \to \bR$, $d_2:\uSpace\times\pSpace \to \bR$ and $d_3:\pSpace\times\pSpace \to \bR$ are given. 
Our discussion can be easily generalized to include pointwise quadratic inequalities, equality constraints, and multiple conditions of each type.

The constraints that define the absorbing set $\absSet$ can be enforced in~\cref{e:restricted-bound} using Lagrange multipliers. Indeed, suppose that if $s \in L^2(\Omega)$ and $\alpha \in \bR$ satisfy $\alpha \geq 0$ and $s(\vec{x}) \geq 0$ almost everywhere on the fluid's domain $\Omega$. Then, inequality~\cref{e:restricted-bound} holds if
\begin{align}
\gamma \geq \,
c(\Lambda \vec{u} - \vec{v}, p) - \ip{\vec{v}}{\vec{f}} - 2\linear(\vec{u}) - \quadratic(\vec{u},\vec{u})
\qquad&
\nonumber\\
+ \alpha [d_0 + d_1(\vec{u},\vec{u}) + d_1(\vec{u},p) + d_3(p,p)] \quad&
\nonumber\\
+ \ip{s}{h + \opK \vec{u} + \opL p} &
\quad \forall (\vec{u},p) \in \uSpace \times \pSpace.
\label{e:s-procedure}
\end{align}
We therefore arrive at the following result.
\begin{theorem}[Background method with extra constraints]\label{th:bm-extra-constraints}
The best upper bound $\gamma^*$ on $\maxAverage$ provable by restricting the background method to the absorbing set $\absSet$ in~\cref{e:absSet} is
\begin{equation}\label{e:bound-estimate-augmented}
	\gamma^* =  \inf_{\substack{\alpha \in \bR,\; \Lambda \in \energyPreservingSpace\\ \vec{v} \in \vSpace,\; s \in L^2(\Omega)}} 
	\left\{ \gamma: \; 
	\text{\cref{e:s-procedure}}, \;
	\alpha \geq 0, \;
	s(\vec{x}) \geq 0 \text{ a.e. on } \Omega
	\right\}.
\end{equation}
\end{theorem}

\begin{remark}
Since~\cref{e:bound-estimate-augmented} reduces to~\eqref{e:bound-sdp-form} when $s(\vec{x})=0$ and $\alpha=0$, the upper bound on $\maxAverage$ in \cref{th:bm-extra-constraints} is always at least as good as that in \cref{th:optimal-bm-general}. However, it can be strictly better; this happens, for instance, in convection driven by uniform internal heating~\cite{Arslan2021} and in RB convection between stress-free boundaries~\cite{Tilgner2019}.
\end{remark}

\begin{remark}
\Cref{th:symmetry-reduction} on symmetry reduction extends to the case in which the full space $\uSpace \times \pSpace$ is replaced by the absorbing set $\absSet$ if the latter is closed under the symmetry group $\group$, i.e., $(\vec{u},p) \in \absSet$ implies that $\GroupAction(\vec{u},p) \in \absSet$ for all $\GroupAction \in \group$. A sufficient condition for this is that the operators $\opK, \opL$ and the bilinear forms $d_1,d_2,d_3$ in~\cref{e:absSet} be invariant under $\group$.
\end{remark}

\begin{figure}
	\centering
	\includegraphics[width=0.95\linewidth]{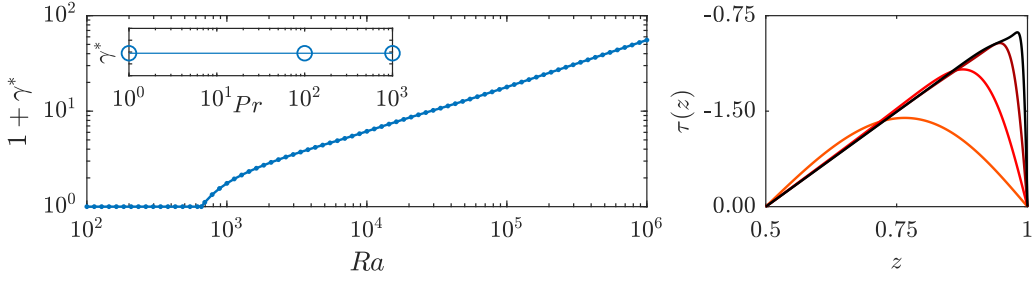}
	\vskip -1ex
	\caption{
		\textit{Left:} Optimal bounds $\gamma^*$ for a two-dimensional version of the RB convection problem in \cref{ex:rb-additional-constraints}  with horizontal half-period $L_x=\sqrt{2}$. These bounds were obtained for fixed combinations of Rayleigh numbers $\Ra$ and Prandtl numbers $\Pran$ by minimizing the value of $\gamma$ for which inequality~\cref{e:rb-augmented-inequality} holds using the methods described in~\cref{ss:sdp-method}. The results are shifted by 1 to ease the visualization and are independent of $\Pran$, as demonstrated in the inset for $\Ra=10^5$.
		\textit{Right:} Optimal generalized background temperature fields at $\Ra = 10^3$ (orange), $10^4$ (red), $10^5$ (brown) and $10^6$ (black). Only half of the domain is shown because all profiles are antisymmetric about $z=\frac12$.
	}
	\label{f:rb-results}
\end{figure}

\begin{example}\label{ex:rb-additional-constraints}
Let us reconsider the problem of RB convection between stress-free boundaries discussed in \cref{ex:rb-bm-general,ex:rb-symmetry}. By the maximum principle, the temperature perturbation $\theta$ satisfies $z-\theta \geq 0$ and $1-z+\theta \geq 0$ at all times if the initial condition also does so, and it approaches such a state exponentially quickly otherwise~\cite{Foias1987}. Moreover, analysis in~\cite{Tilgner2019} shows that
\begin{equation}\label{e:tiglner-estimates}
	\underbrace{\int_{\Omega} \left( 
		\tfrac14 \abs{\nabla \vec{w}}^2 
		\pm \abs{w_3}^2 \pm \Pr \Ra \abs{\partial_z \nabla \Delta^{-1} \theta}^2
		\right)\dVolume}_{=d_1(\vec{u},\vec{u})}
	\underbrace{\pm \int_{\Omega} p \,\partial_z \theta \, \dVolume}_{=d_2(\vec{u},p)} \geq 0
\end{equation}
pointwise in time. The absorbing set $\absSet$ defined by these constraints, which are specific to stress-free boundaries and have the form considered in~\cref{e:absSet}, is invariant under horizontal translation and the ``flow reversal'' operation~\cref{e:rb-flow-reversal}. Thus, an extension of \cref{th:symmetry-reduction} guarantees that there is no loss of generality in taking a zero generalized background velocity field ($\vecphi = 0$) and a generalized background temperature field $\tau = \tau(z)$ depending only on the vertical coordinate. With nonnegative Lagrange multipliers $s_1,s_2 \in L^2(\Omega)$ for the two inequalities on $\theta$ and nonnegative scalar Lagrange multipliers $\alpha_1,\alpha_2$ for the two inequalities in~\cref{e:tiglner-estimates}, condition~\cref{e:s-procedure} for the RB problem then reads
\begin{align}\label{e:rb-augmented-inequality}
	\gamma \geq 
	\int_{\Omega}\left(
	\left(\tfrac{\alpha_1}{4} + \tfrac{\alpha_2}{4} - \tfrac{\lambda_1}{\Ra}\right) \abs{\nabla \vec{w}}^2
	+(\alpha_1 - \alpha_2 + \lambda_3) \abs{w_3}^2
	-\lambda_2 \abs{\nabla \theta}^2 
	\right) \,&\dVolume
	\nonumber\\
	+ \int_{\Omega} \left( ( 1 + \lambda_1 + \lambda_2 - \tau') w_3\theta
	-\lambda_3	(\Pran+1)\nabla\theta \cdot \nabla w_3
	\right) \,&\dVolume
	\nonumber\\
	+\int_{\Omega}\left(
	(\alpha_1 - \alpha_2) \Pran\Ra  \abs{\partial_z \nabla \Delta^{-1} \theta}^2
	+\lambda_3 \Pran\Ra \abs{\theta}^2
	\right) \,&\dVolume
	\nonumber\\
	+ \int_{\Omega} \left(\tau' \partial_z\theta - \tau w_3 + (z-\theta)s_1 + (1-z+\theta)s_2 \right) \,&\dVolume
	\nonumber\\
	+\int_{\Omega} (\alpha_1 - \alpha_2 + \lambda_3) \,p\, \partial_z \theta  \;&\dVolume,
\end{align}
%
where primes denote total derivatives in the $z$ coordinate. Since this inequality depends linearly on the pressure $p$ and must hold for all $p$ in the linear space $\pSpace$, one is forced to choose $\lambda_3 = \alpha_2 - \alpha_1$ so the pressure drops out of the problem. In contrast to the basic formulation in \cref{ex:rb-bm-general}, however, one is \textit{not} forced to set $\lambda_3=0$, which in principle could result in better $\Pran$-dependent bounds on the mean vertical heat flux. Unfortunately, this appears not to be the case in practice: computations using the methods described in~\cref{ss:sdp-method} for a two-dimensional version of the problem, obtained by dropping the $y$ direction, yield the bounds and generalized background temperature fields shown in \cref{f:rb-results} at all values of $\Pran$ we tested. However, the same approach yields $\Pran$-dependent bounds on the mean poloidal kinetic energy~\cite{Tilgner2019}.
\end{example}

\section{Computational implementation}\label{s:implementation}

Calculating the best upper bound on $\maxAverage$ that the background method has to offer requires solving the minimization problem~\cref{e:bound-sdp-form} or its improved version~\cref{e:bound-estimate-augmented}. Usually, these problems are analytically intractable and must be tackled numerically.

The computational implementation of the background method is complicated by the fact that the Euler--Lagrange equations for the optimal $\Lambda$ and $\vec{v}$ in~\cref{e:bound-sdp-form} and for the optimal $\vec{u}$ in~\cref{e:bound-saddle-point-form} are nonlinear and, often, admit multiple solutions. Only solutions satisfying the spectral constraint~\cref{e:spectral-constraint} correspond to the saddle point of problem~\cref{e:bound-saddle-point-form} and give the optimal bound on $\maxAverage$; the others are ``spurious'' stationary points for which the eigenvalue problem~\cref{e:general-eig-prob-sc} has negative eigenvalues (see~\cite[Figure~2]{Wen2015a} for a schematic illustration), and which therefore do not produce valid bounds.

This issue has historically been addressed using delicate numerical schemes based on continuation and bifurcation analysis~\cite{Doering1997,Nicodemus1997a,Nicodemus1998a,Plasting2003,Otero2004,Tang2004,Plasting2005,Wittenberg2010a}. Recently, two simpler alternatives have emerged that do not require continuation and have been applied successfully to a variety of flows.
One approach~\cite{Fantuzzi2015,Fantuzzi2016CDC,Fantuzzi2016PRE,Fantuzzi2017b,Tilgner2017,Fantuzzi2018a,Tilgner2019,Arslan2021} discretizes the minimization problem~\cref{e:bound-sdp-form} into a semidefinite program (SDP)---a convex optimization problem where matrices with affine dependence on the optimization variables are constrained to be positive semidefinite. This SDP can then be solved using algorithms with polynomial-time complexity~\cite{Nesterov1994,Vandenberghe1996,Nemirovski2006,ODonoghue2016,Wen2010}. 
The other approach~\cite{Gambill2006,Wen2013,Wen2015a,Wen2018,Ding2019,Lee2019} applies the steepest gradient method to the saddle-point formulation~\cref{e:bound-saddle-point-form} to derive time-dependent versions of the Euler--Lagrange equations, which can be timestepped until convergence to a stationary solution. 
The next subsections review these two approaches and reveal a new connection between them.

\subsection{Optimal bounds with semidefinite programming}
\label{ss:sdp-method}

To discretize problem \cref{e:bound-sdp-form} into an SDP, suppose that all balance parameters in the linear space $\energyPreservingSpace_0$ can be represented as
\begin{equation}
\Lambda = \sum_{i=1}^s \lambda_i \Lambda_i
\end{equation}
for fixed symmetric matrices $\Lambda_1,\ldots,\Lambda_s$ and some vector of coefficients $\veclambda = (\lambda_1,\ldots,\lambda_s)$. Further, replace the spaces $\vSpace$, $\uSpace$ and $\pSpace$ with finite-dimensional subspaces
\begin{subequations}\label{e:discrete-spaces}
\begin{align}
	\vSpace_h &:= \linspan( \vecxi_1, \ldots, \vecxi_{m_v}), \\
	\uSpace_h &:= \linspan( \veceta_1, \ldots, \veceta_{m_u}), \\
	\pSpace_h &:= \linspan( \zeta_1, \ldots, \zeta_{m_p}),
\end{align}
\end{subequations}
where $\vecxi_i$, $\veceta_i$ and $\zeta_i$ are prescribed basis functions and $h$ is a discretization parameter inversely proportional to the dimension of the discrete spaces (e.g., $h$ represents the size of a mesh for the fluid's domain). We assume for simplicity that $\vecxi_i$ and $\veceta_i$ satisfy all constraints used to define $\vSpace$ and $\uSpace$, such as boundary conditions and incompressibility; other choices may be more convenient in practice, but require enforcing these conditions using projection operators or Lagrange multipliers that complicate the exposition. 

Expanding 
\begin{align}
\vec{v}(\vec{x}) &= \sum_{i=1}^{m_v} \hat{v}_i \,\vecxi_i(\vec{x}), &
\vec{u}(\vec{x}) &= \sum_{i=1}^{m_u} \hat{u}_i \veceta_i(\vec{x}), &
p(\vec{x}) &= \sum_{i=1}^{m_p} \hat{p}_i \,\zeta_i(\vec{x}),
\end{align}
and writing 
$\hat{\vec{v}} = (\hat{v}_1,\ldots,\hat{v}_{m_v})$ and $\hat{\vec{u}} = (\hat{u}_1,\ldots,\hat{u}_{m_u})$ for the vectors of expansion coefficients, the inequality constraint in~\cref{e:bound-sdp-form} discretizes into
\begin{equation}\label{e:lmi}
\begin{pmatrix}1 \\ \hat{\vec{u}}\end{pmatrix}^\tr
\underbrace{
	\begin{pmatrix}
		\gamma + \vec{f}^h \cdot \hat{\vec{v}} & \vec{l}(\veclambda,\hat{\vec{v}})^\tr\\
		\vec{l}(\veclambda,\hat{\vec{v}}) & \tfrac12 Q(\veclambda,\hat{\vec{v}})  + \tfrac12 Q(\veclambda,\hat{\vec{v}}) ^\tr
	\end{pmatrix}
}_{=:S(\gamma, \veclambda, \hat{\vec{v}})}
\begin{pmatrix}1 \\ \hat{\vec{u}}\end{pmatrix}
\geq 0 \quad \forall \hat{\vec{u}} \in \bR^{m_u},
\end{equation}
where the (column) vectors $\vec{f}^h$ and $\vec{l}$ and the matrix $Q$ have entries
\begin{subequations}
\begin{gather}
	f^h_i := \ip{\vec{f}}{\vecxi_i}, \quad i = 1,\ldots,m_v,\\
	{l}_i := 
	\frac12 \sum_{j=1}^{m_v} a(\vecxi_j, \veceta_i)\, \hat{v}_j 
	-\frac12 \sum_{j=1}^s \ip{\vec{f}}{\Lambda_j \veceta_i} \lambda_j,
	\quad i = 1,\ldots,m_u,\\
	Q_{ij} :=
	\sum_{k=1}^{m_v} b(\vecxi_k, \veceta_i, \veceta_j) \hat{v}_k
	- \sum_{k=1}^s a(\Lambda_k \veceta_i, \veceta_j) \lambda_k 
	- \Phi(\veceta_i,\veceta_j),
	\quad i,j = 1,\ldots,m_u.
\end{gather}
\end{subequations}
It is well known that~\cref{e:lmi} holds if and only if the matrix $S(\gamma, \veclambda, \hat{\vec{v}})$ is positive semidefinite, a \textit{linear matrix inequality} (LMI) constraint which we denoted by $S(\gamma, \veclambda, \hat{\vec{v}})\succeq 0$. The constraint $\opC^\star \vec{v} = 0$, instead, can be imposed in a weak sense by requiring $c(\vec{v},p) = 0$ for all $p \in \pSpace_h$, leading to the linear constraints $C^\tr \hat{\vec{v}} = 0$ with matrix coefficients  $C_{ij} := c(\vecxi_i, \zeta_j)$ for $i=1,\ldots,m_v$ and $j = 1,\ldots,m_p$. Thus, one obtains the SDP
\begin{equation}\label{e:sdp}
\gamma_h^* = \inf_{\substack{\lambda \in \bR^s \\ \hat{\vec{v}} \in \bR^{m_v}}} \;
\left\{ \gamma: \;S(\gamma, \veclambda, \hat{\vec{v}}) \succeq 0 \text{ and } C^\tr \hat{\vec{v}} = 0 \right\}.
\end{equation}
%
If the discrete spaces $\vSpace_h$, $\uSpace_h$ and $\pSpace_h$ enjoy suitable approximation properties in $\vSpace$, $\uSpace$ and $\pSpace$, it is reasonable to expect that the optimal solution $\gamma_h^*$ of this SDP converges to the optimal bound $\gamma^*$ in \cref{th:optimal-bm-general} as the discretization is refined ($h \to 0$). To the best of our knowledge, however, rigorous convergence results are yet to be proven.

The SDP~\cref{e:sdp} can be solved using a variety of general-purpose algorithms~\cite{Nesterov1994,Vandenberghe1996,Nemirovski2006,ODonoghue2016,Wen2010}, which can be divided into the two categories of \textit{interior-point methods} and \textit{first-order methods}. The former are versions of Newton's method in which positive semidefinite constraints are imposed via so-called barrier functions, and converge to $\varepsilon$-suboptimal solutions in $O(\abs{\log\varepsilon})$ iterations~\cite[\S4.3.3]{Nesterov2003} (often, no more than 10--50). However, for an SDP with an $n\times n$ LMI and $m$ optimization variables, each iteration of a standard interior-point method requires $O(n^2+m^2)$ memory and $O(m^3+n^2m^2+n^3m)$ floating-point operations~\cite[\S4.3.3]{Nesterov2003}, both of which increase quickly as $n$ or $m$ are raised. First-order methods, instead, have a lower complexity because they use only gradient information, but also converge more slowly. For example, the algorithm in~\cite{Wen2010} finds  $\varepsilon$-suboptimal solutions in $O(\frac1\varepsilon)$ iterations that require $O(m^2 + n^3)$ floating-point operations each.

It must be observed that the solution $\gamma_h^*$ of~\cref{e:sdp} is not a rigorous upper bound on $\maxAverage$, because replacing the spaces $\uSpace$ and $\pSpace$ with their subspaces $\uSpace_h$ and $\pSpace_h$ relaxes the constraint in the original problem~\cref{e:bound-sdp-form}. On the other hand, one can always optimize $\vec{v}$ over the finite-dimensional subspace $\vSpace_h$ of $\vSpace$ at the expense of worsening the bound on $\maxAverage$. Consequently, $\gamma_h^*$ estimates from \textit{below} the best (suboptimal) upper bound on $\maxAverage$ that can be proven with $\vec{v} \in \vSpace_h$. To obtain an estimate from above, one must modify the constraints of~\cref{e:sdp} to account for the difference between $\uSpace$ and $\pSpace$ and their discrete counterparts. This is achieved in~\cite{Fantuzzi2016PRE,Fantuzzi2016CDC,Fantuzzi2017b} for discretizations based on Legendre expansions in one coordinate and Fourier expansions in the remaining ones: using estimates that rely on the special properties of the Fourier and Legendre bases, such as orthogonality and differentiation rules, the infinite-dimensional constraint in~\cref{e:bound-sdp-form} can be  \textit{strengthened} into sufficient finite-dimensional LMIs, which depend on a finite set of expansion coefficients and on the $L^2$ norm of the expansion `tail'. Generalizing these results to other expansion bases remains an open problem.

\begin{figure}
	\centering
	\includegraphics[height=4.5cm]{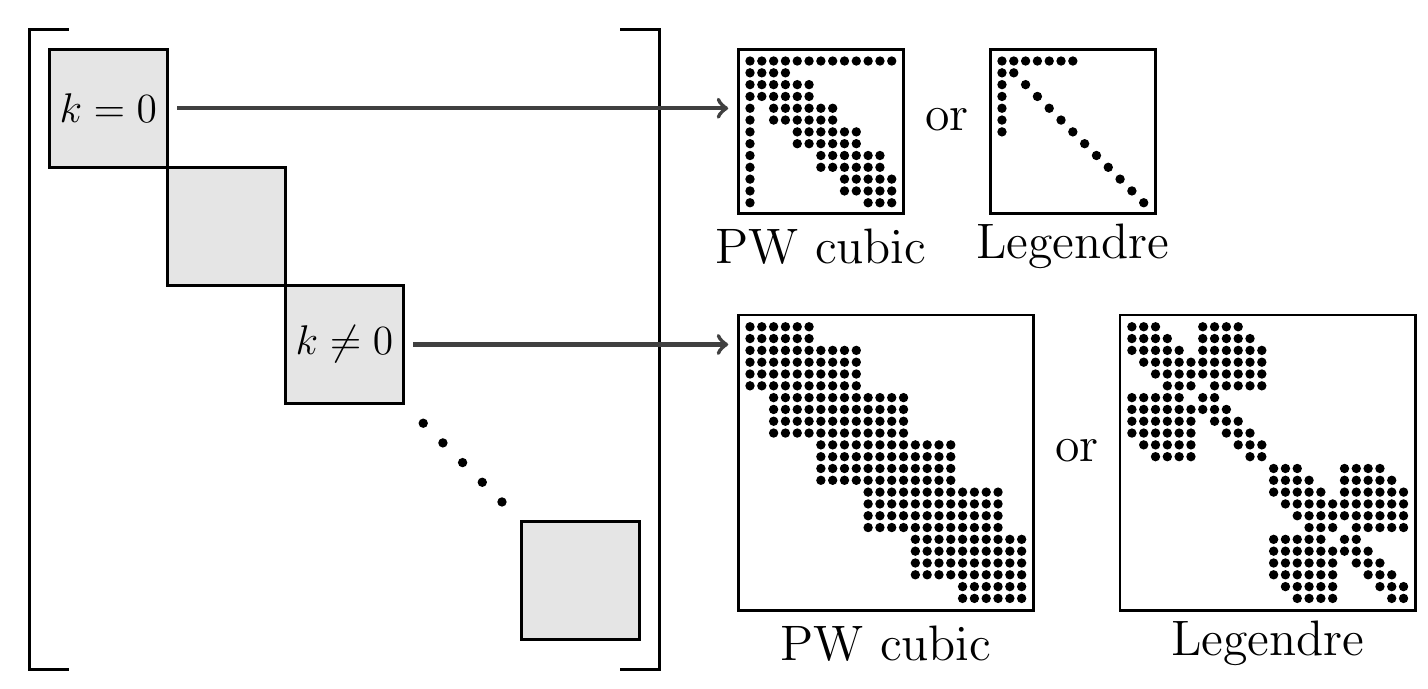}\\[10pt]
	\includegraphics[width=0.99\textwidth]{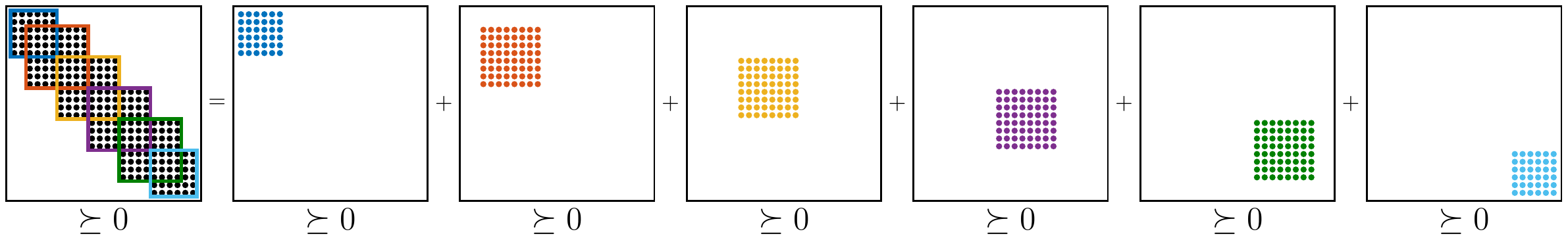}
	\vskip -1ex
	\caption{
		\textit{Top:} Structure of the LMI~\cref{e:lmi} for a two-dimensional version of the RB convection problem in \cref{ex:rb-bm-general} with $\lambda_3=0$ and after symmetry reduction. The basis functions are products of sinusoidal functions with wavenumber $k$ in the horizontal direction, and either Legendre polynomials or piecewise (PW) cubic basis functions with compact support in the vertical direction. \textit{Bottom:} Chordal decomposition of a block with nonzero wavenumber $k$ for the PW cubic case.
	}
	\label{fig:sparsity_matrices}
\end{figure}

Another crucial observation is that the choice of basis functions for $\vSpace_h$, $\uSpace_h$ and $\pSpace_h$ strongly influences the structure of the LMI constraint in~\cref{e:sdp} and, therefore, its computational complexity. For instance, \cref{fig:sparsity_matrices} illustrates different LMI structures obtained when a two-dimensional version of the RB problem considered in \cref{ex:rb-bm-general}, simplified as described in \cref{ex:rb-symmetry}, is discretized using sinusoidal functions with wavenumber ${k}$ in the horizontal direction and either Legendre polynomials or compactly-supported piecewise-cubic functions in the vertical direction. The block-diagonal structure corresponds to the decoupling of different wavenumbers, and the positive semidefiniteness of each block can be imposed separately to obtain an SDP with multiple smaller LMIs. This is convenient because SDPs of this type can currently be solved more efficiently than SDPs with a single large LMI; for instance, standard interior point methods require $O(m^3+ks^2m^2 + ks^3m)$ floating-point operations per iteration to handle a $ks \times ks$ block-diagonal LMI with $m$ optimization variable and $k$ diagonal blocks of size $s \times s$, which is usually much smaller than the $O(m^3+k^2s^2m^2 + k^3s^3m)$ operations required when the block-diagonal structure is not exploited.

\begin{figure}
	\centering
	\includegraphics[width=0.9\textwidth]{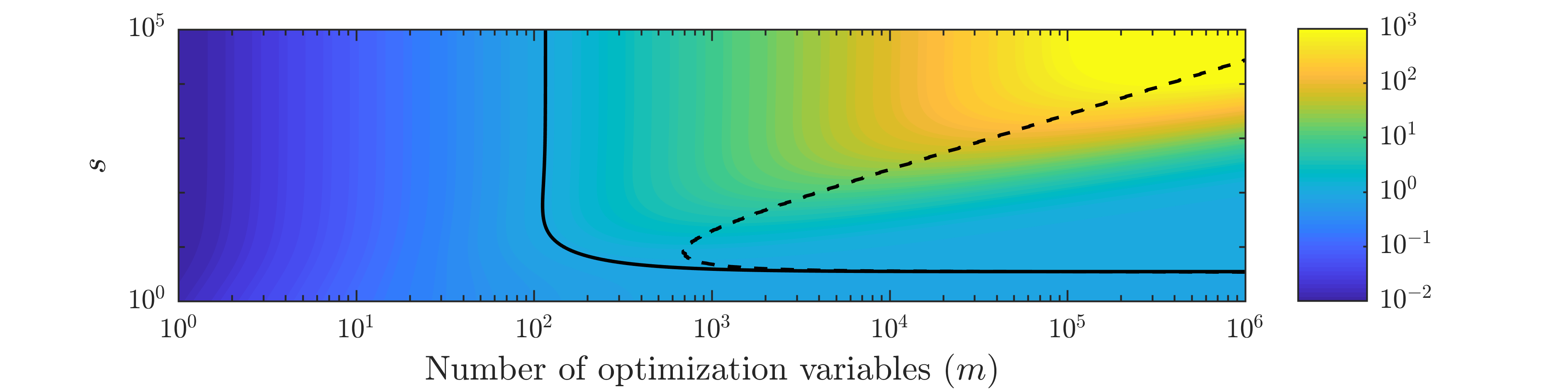}
	\vskip -1ex
	\caption{
		Ratio between the floating-point operations required by standard interior point methods before and after the chordal decomposition of an LMI with the structure illustrated in the bottom panel of \cref{fig:sparsity_matrices}. Here, $m$ is the number of optimization variable, and $s$ is the number of $8 \times 8$ partially overlapping blocks ($s=4$ in \cref{fig:sparsity_matrices}). The size of the LMI is $(4s+8) \times (4s \times 8)$. Large contour values indicate large computational savings due to chordal decomposition. To the left of the solid line (small $s$ or $m$), chordal decomposition brings no savings. The dashed line marks values of $m$ for which, given $s$, chordal decomposition brings maximum advantage.
	}
	\label{fig:flop-improvement}
\end{figure}

The structure of each block can be exploited in a similar way using \textit{chordal decomposition} techniques for SDPs~\cite{Fukuda2000,Nakata2003,Vandenberghe2015,Zheng2019a,ZhengFantuzzi2021review}, which decompose sparse LMIs into smaller ones by considering their dense principal submatrices (see the bottom panel in \cref{fig:sparsity_matrices} for an illustration). This requires introducing additional optimization variables to account for the overlap between dense submatrices, but, if these are small and do not overlap significantly, then the added cost is negligible compared to the savings associated by the reduction in LMI dimension. As illustrated in \cref{fig:flop-improvement}, these savings can be significant for general LMIs with the structure shown in the bottom panel of \cref{fig:sparsity_matrices} ($s$ is the number of partially overlapping $8\times 8$ blocks, equal to $4$ in \cref{fig:sparsity_matrices}, while $m$ is the number of optimization variables). One can therefore optimize generalized background fields with very sharp and possibly nonsmooth boundary layers~\cite{Fantuzzi2018a,Arslan2021}.
However, it must be kept in mind that a decomposition into dense submatrices does not strengthen the LMI in~\cref{e:sdp} only when its structure satisfies the following technical condition, which holds for the piecewise-cubic case in \cref{fig:sparsity_matrices} but not for the case of Legendre polynomials: for every integer $\alpha \geq 3$, if the entries of $S(\gamma,\veclambda,\hat{\vec{v}})$ in position $(i_1,i_2)$, $(i_2,i_3)$, $\ldots$, $(i_{\alpha-1},i_\alpha)$ and $(i_\alpha, i_1)$ are nonzero, then there exist $\beta,\gamma \in \{1,\ldots,\alpha\}$ with $\gamma> \beta+1$ such that the entry in position $(i_\beta,i_\gamma)$ is also nonzero. Employing matrix decomposition when this condition does not hold is restrictive and, in general, leads to conservative bounds on $\maxAverage$. It remains to be determined whether the restriction is or is not mild enough for these conservative bounds to be useful.

Finally, the improved background method formulation outlined in \cref{ss:additional-constraints} can be implemented using semidefinite programming following a very similar discretization strategy. However, care must be taken to enforce the nonnegativity of any space-dependent Lagrange multipliers exactly. The task can be simplified considerably by a careful choice of the basis functions used to discretize these Lagrange multipliers; piecewise-linear basis functions on a finite-element mesh, for instance, lead to simple linear inequalities on the values of the Lagrange multipliers on the nodes of the mesh~\cite{Fantuzzi2018a,Arslan2021}.

\subsection{Optimal bounds with timestepping}
\label{ss:timestepping}

An alternative approach, used to optimize bounds for the Kuramoto--Sivashinsky equation in~\cite{Gambill2006} and extended  to fluid flows in~\cite{Wen2013,Wen2015a}, is to view the optimization variables in the saddle-point problem~\cref{e:bound-saddle-point-form} as time-dependent and derive evolution equations ensuring the steepest ascent in $\vec{u}$ and the steepest descent in $\vec{v}$ and $\Lambda$. Well-known and efficient numerical schemes for differential equations can then be employed to solve such equations until convergence to a steady solution, which is a critical point for~\cref{e:bound-saddle-point-form} and, very often, is the correct saddle point. This approach has been shown to work robustly for Rayleigh--B\'enard convection~\cite{Wen2015a}, porous media convection~\cite{Wen2018}, plane Couette flow with injection and suction~\cite{Lee2019}, and Taylor--Couette flow~\cite{Ding2019}.

A subtle but essential observation (see, e.g.,~\cite[Ch.~3]{Fantuzzi2018b} and~\cite{Ding2020}) is that tracking the time evolution of a single field $\vec{u} \in \uSpace$ may not be sufficient to solve~\cref{e:bound-saddle-point-form} correctly. This is because, when the balance parameter  $\Lambda$ and the generalized background field $\mathbf{v}$ are optimal, the quadratic form $\quadratic$ may vanish for a number $\rho$ of fields $\vec{u}_1,\ldots,\vec{u}_\rho \in \uSpace$, which are often called \textit{critical modes}~\cite{Plasting2003,Wen2013,Wen2015a} and are eigenfunctions with zero eigenvalue for the eigenvalue problem~\cref{e:general-eig-prob-sc} associated to the spectral constraint. For reasons that will become clear in~\cref{ss:links} these critical fields must be tracked explicitly to impose the spectral constraint, even though they give no contribution to the optimal value of~\cref{e:bound-saddle-point-form} by \cref{prop:spectral-constraint}\cref{sc:compatibility}. One must also track a generic field, denoted by $\mathbf{u}_0$, for which $q$ need not vanish and which plays the role of $\vec{u}$ in~\cref{e:bound-saddle-point-form}. In other words, rather than considering the objective functional in~\cref{e:bound-saddle-point-form}, one must look for a saddle point of the modified Lagrangian
\begin{equation}\label{e:timestepping-Lagrangian}
L(\vec{u}_0,\ldots,\vec{u}_\rho,p,\vec{v},\Lambda) := -c(\vec{v},p) -\ip{\vec{v}}{\vec{f}} - 2 \linear(\vec{u}_0) - \sum_{r=0}^\rho \quadratic(\vec{u}_r, \vec{u}_r),
\end{equation}
where $p$ is a Lagrange multiplier enforcing the constraint $\opC^\star \vec{v}=0$. 

Writing $\Lambda = \sum_{j=1}^s \lambda_j  \Lambda_j$ as in~\cref{ss:sdp-method}, the time derivative of this Lagrangian is
\begin{equation}
\ddt L =
\bigip {\frac{\delta L}{ \delta \vec{v} } }{ \ddt{\vec{v}} } + 
\sum_{j=1}^s \bigip{\frac{\partial L}{\partial \lambda_j}}{ \ddt{ \lambda_j}} +
\sum_{r=0}^\rho \bigip{\frac{\delta L}{\delta \vec{u}_r}}{ \ddt{\vec{u}_r} }
-\bigip{\opC^\star \vec{v}}{ \ddt p},
\end{equation}
where $\frac{\delta L}{\delta \psi}$ denotes the variation (Fr\'echet derivative) of $L$ with respect to a field $\psi$. Insisting that $\opC^\star \vec{v}=0$ pointwise in time, so the last term vanishes, the evolution equations ensuring the fastest growth in $L$ with respect to each $\vec{u}_r$ and the fastest reduction with respect to $\vec{v}$ and each $\lambda_j$ are
\begin{subequations}
\label{e:time-marching}
\begin{align}
	\label{e:time-marching-v}
	&&\ddt{\vec{v}} &= - \frac{\delta L}{\delta \vec{v}} \equiv \opA \vec{u}_0 + \sum_{r=0}^\rho \opB(\vec{u}_r,\vec{u}_r) + \opC p + \vec{f},
	&&\\
	\label{e:time-marching-lambda}
	&&\ddt{\lambda_j} &= - \frac{\partial L}{\partial \lambda_j}  \equiv - \ip{\vec{f}}{\Lambda_j \vec{u}_0} - \sum_{r=0}^\rho a(\Lambda_j\vec{u}_r,\vec{u}_r)
	&&\text{for } j =1,\ldots,s,&&
	\\
	\label{e:time-marching-u}
	&&\ddt{\vec{u}_r} &= \;\frac{\delta L}{\delta \vec{u}_r} \; \equiv \bigg( \sum_{j=1}^s \lambda_j \Lambda_j \vec{f} - \opA^\star \vec{v} \bigg) \delta_{0r} - \opF(\vec{v}) \vec{u}_r
	&&\text{for }r =0,\ldots,\rho,&&
\end{align}
\end{subequations}
where $\delta_{0r}$ is the usual Kronecker delta and we have introduced the linear operator $\opF(\vec{v}) := \opH(\vec{v}) + \opH(\vec{v})^\star - \Lambda \opA - \opA^\star \Lambda - \opP - \opP^\star$ for notational convenience. Stationary solutions to these equations are stationary points of the Lagrangian $L$, and those satisfying the spectral constraint~\cref{e:spectral-constraint} provide optimal solutions to~\cref{e:bound-saddle-point-form}.

Projecting~\cref{e:time-marching-v,,e:time-marching-lambda,,e:time-marching-u} onto the finite-dimensional spaces $\uSpace_h$, $\vSpace_h$ and $\pSpace_h$ defined in~\cref{e:discrete-spaces} yields the set of ODEs
\begin{subequations}
\label{e:discrete-time-marching}
\begin{align}
	\ddt{(V\hat{\vec{v}})} &= A \hat{\vec{u}}_0 + \sum_{r=0}^\rho \vec{b}(\hat{\vec{u}}_r,\hat{\vec{u}}_r) + C\hat{\vec{p}} + \vec{f}^h,
	\\
	\ddt{\lambda_j} &= - \sum_{i=1}^{m_u} \ip{\vec{f}}{\Lambda_j \veceta_i} \hat{u}_{0i} 
	- \sum_{r=0}^{\rho} \, \sum_{i,k=1}^{m_u} \! a(\Lambda_j \veceta_i, \veceta_k) \hat{u}_{ri} \hat{u}_{rk}
	&&\text{for }j =1,\ldots,s,
	\\
	\ddt{(U \hat{\vec{u}}_r)} &= -2\vec{l}(\veclambda, \hat{\vec{v}}) \delta_{0r} - \left[ Q(\veclambda, \hat{\vec{v}}) + Q(\veclambda, \hat{\vec{v}})^\tr\right] \hat{\vec{u}}_r, \phantom{\sum_{r=0}^{\rho}}
	&&\text{for }r =0,\ldots,\rho.
\end{align}
\end{subequations}
Here, $\hat{\vec{u}}_r = (\hat{u}_{r1},\ldots,\hat{u}_{rm_u})$ is the vector of expansion coefficients of $\vec{u}_r$ in the chosen basis for $\uSpace_h$, the vector $\vec{b}(\hat{\vec{u}}_r,\hat{\vec{u}}_r)$ has entries
\begin{equation}
b_i(\hat{\vec{u}}_r,\hat{\vec{u}}_r) := \sum_{j,k=1}^{m_u} b(\vecxi_i,\veceta_j,\veceta_k) \hat{u}_{rj} \, \hat{u}_{rk}, \qquad i=1,\ldots,m_v,
\end{equation}
the matrices $V$, $A$ and $U$ have entries
\begin{subequations}
\begin{align}
	&&&&V_{ij} &= \ip{\vecxi_i}{\vecxi_j}, & i,j&=1,\ldots,m_v, &&&&\\
	&&&&A_{ij} &= a(\vecxi_i,\veceta_j), & i&=1,\ldots,m_v,\; j=1,\ldots,m_u,\\
	&&&&U_{ij} &= \ip{\veceta_i}{\veceta_j}, &i,j&=1,\ldots,m_u,
\end{align}
\end{subequations}
and all other quantities are defined in~\cref{ss:sdp-method}.
The ODEs in~\cref{e:discrete-time-marching}, complemented by the algebraic condition $C^\tr \hat{\vec{v}}=0$ obtained upon discretizing the constraint $\opC^\star \vec{v}=0$ as explained in~\cref{ss:sdp-method}, can be integrated in time until convergence to a stationary solution using standard numerical schemes, starting from any randomly generated nonzero initial condition.

\begin{remark}
The value of $\rho$ in~\cref{e:timestepping-Lagrangian} must be chosen \textit{a priori} and must be large enough to ensure that the saddle point of $L$ is the only stationary attractor for~\cref{e:time-marching}, so the timestepping process cannot converge to a ``spurious'' solution. The discussion in~\cref{ss:links} below and the numerical results in~\cite{Wen2013,Wen2015a,Wen2018,Ding2019,Lee2019} and~\cite[Ch.~3]{Fantuzzi2018b} lead us to conjecture that $\rho$ should be equal to the largest possible multiplicity of the principal eigenvalue of the eigenvalue problem~\cref{e:general-eig-prob-sc} associated to the spectral constraint~\cref{e:spectral-constraint}. Analysis in~\cite{Wen2015a} establishes this conjecture when the background method is applied with \textit{fixed} balance parameter $\Lambda$ (i.e., fixed $\lambda_1,\ldots,\lambda_s$) to RB convection with stress-free isothermal boundaries, plane Couette flow, and porous media convection. It remains an open theoretical problem to extend these specialized arguments to the general setting considered in the present work.
\end{remark}

\begin{remark}\label{remark:small-timesteps}
In highly turbulent regimes (e.g., convection at very high Rayleigh number), the numerical solution of the ODE system~\cref{e:discrete-time-marching} typically requires small timesteps to ensure numerical stability. In such regimes, therefore, convergence to a stationary solution can be slow. One way to mitigate this issue is to employ a two-step approach~\cite{Wen2015a} in which the timestepping is stopped as soon as the magnitude of the time derivatives drops below a moderate tolerance, and the resulting approximately stationary solution is used to initialize Newton--Kantorovich iterations for the Euler--Lagrange equations of the saddle-point problem~\cref{e:timestepping-Lagrangian}. As discussed in~\cite{Wen2013,Wen2015a}, efficiency can be further improved by adaptively dropping from the computation any vectors $\hat{\vec{u}}_r$ with $r \geq 1$ that appear to converge to zero, because they have no effect on the long-time behaviour of solutions to~\cref{e:discrete-time-marching}.
\end{remark}

\subsection{Connecting the two approaches}
\label{ss:links}

The timestepping strategy of~\cref{ss:timestepping} is closely related to the semidefinite programming approach of~\cref{ss:sdp-method}, in the sense that it provides a particular numerical scheme for solving the SDP~\cref{e:sdp}. Precisely, the ODEs in \cref{e:discrete-time-marching} can be viewed as a steepest-gradient algorithm applied to the SDP~\cref{e:sdp} when the matrix Lagrange multiplier for its LMI constraint is represented using a low-rank factorization.

To make this interpretation evident, let us assume that the SDP~\cref{e:sdp} is strictly feasible, meaning that there exist $\veclambda$ and $\hat{\vec{v}}$ such that $C^\tr \hat{\vec{v}}=0$ and that the matrix $S(\gamma, \veclambda,\hat{\vec{v}})$ is strictly positive definite. With this additional (mild) assumption, standard duality theory for SDPs~\cite[\S5.9.1]{Boyd2004} guarantees that
\begin{equation}\label{e:discrete-lagrangian-orig}
\gamma_h^* = 
\mathop{\vphantom{p}\inf}\limits_{\substack{ \gamma \in \bR,\; \hat{\vec{v}} \in \bR^{m_v} \\ \veclambda \in \bR^s }} \;
\sup_{ \substack{\hat{p} \in \bR^{m_p} \\ Z \succeq 0} } \;
\left\{ \gamma - \hat{\vec{v}} \cdot C \hat{\vec{p}} - Z \cdot S(\gamma, \veclambda,\hat{\vec{v}}) \right\},
\end{equation}
where $\hat{\vec{p}}$ and $Z$ are Lagrange multipliers for the constraints of~\cref{e:sdp}. Moreover, by a condition known as \textit{complementary slackness}, the ranks of the optimal $Z$ and of the optimal $S(\gamma, \veclambda,\hat{\vec{v}})$ must sum to the matrix size $m_u+1$. Therefore, 
if one can anticipate that the optimal solution of the SDP~\cref{e:sdp} satisfies $\rank(S(\gamma, \veclambda,\hat{\vec{v}})) \geq m_u-\rho$ for some integer $\rho\leq m_u$, then the corresponding optimal multiplier must have $\rank(Z)\leq \rho+1$. 

Now, any positive semidefinite matrix $Z$ of dimension $m_u+1$ and rank no larger than $\rho+1$ can be expressed as the sum of at most $\rho+1$ rank-1 matrices. In particular, we can write
\begin{equation}
Z = \begin{pmatrix}t^2 & t\hat{\vec{u}}_0^\tr\\ t\hat{\vec{u}}_0 &  \hat{\vec{u}}_0 \hat{\vec{u}}_0^\tr \end{pmatrix}
+ \sum_{r=1}^\rho \begin{pmatrix}0 & \vec{0}^\tr\\ \vec{0} &  \hat{\vec{u}}_r \hat{\vec{u}}_r^\tr \end{pmatrix}
\end{equation}
for some vectors $\hat{\vec{u}}_0,\ldots,\hat{\vec{u}}_\rho \in \bR^{m_u}$ and some nonnegative scalar $t$. Substituting this decomposition and the definition of $S(\gamma, \veclambda,\hat{\vec{v}})$ from~\cref{e:lmi} into~\cref{e:discrete-lagrangian-orig}, and unfolding matrix inner products involving rank-1 matrices into standard matrix-vector products, leads to
\begin{multline}\label{e:discrete-lagrangian-factorized}
\gamma_h^* = 
\mathop{\vphantom{p}\inf}\limits_{\substack{ \gamma \in \bR \\ \hat{\vec{v}} \in \bR^{m_v} \\ \veclambda \in \bR^s }}\;
\sup_{ \substack{ \hat{\vec{u}}_0,\ldots,\hat{\vec{u}}_\rho \in \bR^{m_u} \\ \hat{p} \in \bR^{m_p} \\ t \geq 0 } }
\bigg\{ (1-t^2)\gamma - \hat{\vec{v}} \cdot C \hat{\vec{p}}
- t^2\, \vec{f}^h \cdot \hat{\vec{v}}
\\[-6ex]
- 2 t \, \vec{l}(\veclambda, \hat{\vec{v}}) \cdot \hat{\vec{u}}_0
- \sum_{r=0}^\rho \hat{\vec{u}}_r \cdot Q(\veclambda,\hat{\vec{v}}) \hat{\vec{u}}_r \bigg\}.
\end{multline}
Optimizing over $\gamma$ yields $t=1$, in which case the argument of this inf-sup problem 
reduces \textit{exactly} to the discrete version of the Lagrangian $L$ in~\cref{e:timestepping-Lagrangian}. Further optimizing over $\hat{\vec{p}}$ returns the constraint $C^\tr \hat{\vec{v}}=0$, while the ODEs in~\cref{e:discrete-time-marching} are recovered upon fixing $t=1$ and applying steepest ascent in each $\hat{\vec{u}}_r$ and steepest descent in $\hat{\vec{v}}$ and $\veclambda$ with respect to the discrete norms 
$\|\hat{\vec{u}}_r\|_{\uSpace_h} = ( \hat{\vec{u}}_r \cdot U \hat{\vec{u}}_r  )^{1/2}$,  
$\|\hat{\vec{v}}\|_{\vSpace_h} = ( \hat{\vec{v}} \cdot V \hat{\vec{v}} )^{1/2}$ and 
$\|\veclambda\|_2 = \sqrt{\veclambda \cdot \veclambda}$. 
At the discrete level, therefore, the timestepping approach of~\cref{ss:timestepping} is a particular example of general low-rank factorization strategies for SDPs that have been studied extensively~\cite{Burer2002,Burer2003,Burer2005,Burer2006,Boumal2020,Waldspurger2020}.

This realization enables us to make interesting observations regarding the theoretical convergence properties of the timestepping approach---specifically, about whether one can guarantee the avoidance of spurious solutions in general, as suggested by the numerical evidence in~\cite{Wen2013,Wen2015a,Wen2018,Ding2019,Lee2019}. On the one hand, the timestepping approach should not be expected to work correctly unless $\rho+1$ is at least as large as the rank of the optimal matrix $Z$ in~\cref{e:discrete-lagrangian-orig}. This is why it is usually necessary to include the fields $\vec{u}_1,\ldots,\vec{u}_\rho$ in~\cref{e:timestepping-Lagrangian}. By complementary slackness, the rank of the optimal $Z$ coincides with the number of zero eigenvalues of the optimal $S(\gamma,\veclambda,\hat{\vec{v}})$. Since the bottom-right block of this matrix is the discrete counterpart of the self-adjoint operator on the left-hand side of the eigenvalue problem~\cref{e:general-eig-prob-sc}, we conclude that $\rho$ should be no smaller than the maximum possible multiplicity of the principal eigenvalue of that problem as $\vec{v}$ and $\Lambda= \sum_{j=1}^s \lambda_j \Lambda_j$ are varied. 

On the other hand, if we assume that the SDP~\cref{e:sdp} and its dual have smooth constraint sets in the sense of~\cite{Boumal2020}, then general results for low-rank factorization methods for SDPs~\cite{Boumal2020} guarantee that the timestepping approach avoids spurious solutions when $\rho+1$ is larger than an upper bound on the rank of the optimal $Z$ derived in~\cite{Pataki1998}, which depends only on the size of the optimization variables in~\cref{e:discrete-lagrangian-orig}. This bound, however, is usually much larger than the maximum eigenvalue multiplicity mentioned above, so its use results in significantly more expensive computations. Unfortunately, it is also sharp~\cite{Waldspurger2020}: there exist SDPs (not necessarily arising from applications of the background method) for which setting $\rho+1$ to be larger than the rank of the optimal $Z$ but smaller than the upper bound from~\cite{Pataki1998} can result in convergence to locally stable spurious solutions. This difficulty, moreover, cannot be resolved by replacing gradient-based timestepping with another algorithm for saddle-point problems because the results in~\cite{Waldspurger2020} are independent of the numerical scheme used to solve~\cref{e:discrete-lagrangian-factorized}. It would be interesting to investigate whether locally stable spurious solutions can arise for SDPs coming from the background method, and one possible way forward is to try and combine the general analysis in~\cite{Boumal2020,Waldspurger2020} with the convergence proofs for specific flows in~\cite{Wen2015a}. Based on the available numerical evidence in~\cite{Wen2013,Wen2015a,Wen2018,Ding2019,Lee2019} and in~\cite[Ch.~3]{Fantuzzi2018b}, we make the following conjecture.

\begin{conjecture*}
Taking $\rho$ equal to the maximum possible multiplicity of the eigenvalue problem~\cref{e:general-eig-prob-sc} suffices to optimize background fields correctly.
\end{conjecture*}

While proving or disproving this statement remains an open theoretical problem, we stress that the timestepping approach works robustly in practice and we are not aware of any examples where locally stable spurious solutions arise once $\rho$ is chosen according to our conjecture. (An example that smaller values can result in convergence to spurious solutions, instead, is given in~\cite[Ch.~3]{Fantuzzi2018b}). Furthermore, in particular implementations one can simply choose $\rho$ as conjectured, check if timestepping returns a spurious solution by solving the eigenvalue problem~\cref{e:general-eig-prob-sc} 
and, in that case, repeat the calculation with a larger~$\rho$. 

\section{Conclusions}\label{s:conclusions}

Despite being established over 30 years ago, the background method remains one of the main tools to place rigorous bounds on time-averaged properties of turbulent flows. Here, we have reviewed a recent interpretation of the method as an application of the auxiliary function framework~\cite{Chernyshenko2014a,Fantuzzi2016b,Chernyshenko2017} with quadratic auxiliary functions in the form~\cref{e:bm-af}, which are parametrized by generalized background fields and balance parameters. This interpretation allows one to systematically formulate convex variational principles to search for bounds on mean quantities that, contrary to the traditional scope of the background method~\cite{Constantin1995}, need not be equivalent to the mean dissipation in the flow. We have also shown that symmetries can be used to simplify the choice of (generalized) background fields---proving, in particular, that nonzero background velocity fields cannot improve existing bounds on the heat transport in Rayleigh--B\'enard convection---and that additional constraints, such as maximum principles, can be enforced using Lagrange multipliers. We hope that our general presentation using the abstract flow description of \cref{s:governing-equations}  can provide a ``recipe'' for future applications of the background method to a wider variety of flows, and offer a basis for future refinements and extensions.

We have also provided a general description of two approaches for the computational implementation of the background method, which have been applied successfully to a variety of shear and convective flows in horizontally periodic domains (see~\cref{table:list_of_addressed_problems} for references). The first approach, based on semidefinite programming, is extremely flexible, has general convergence guarantees, and can be implemented using general-purpose interior-point SDP solvers that are available open-source. To use such solvers in turbulent regimes that require accurate discretization of thin boundary layers, however, one must employ advanced decomposition techniques that have only recently been developed by the optimization community (see~\cite{ZhengFantuzzi2021review} for a review). The second approach to optimizing background fields we have reviewed, instead, relies on timestepping. This is generally more familiar to the fluid mechanics community and can more easily reach highly turbulent regimes, but can converge slowly and presently lacks general theoretical convergence guarantees beyond the particular ones in~\cite{Wen2015a}. \Cref{t:comparison} summarizes the main advantages and drawbacks of the two methods, in the hope of helping readers choose which one to use in particular applications.

\begin{table}
\centering
\caption{Comparison between the two approaches to numerically optimize background fields reviewed in this work. For each property, checkmarks indicate which approach should currently be preferred.}
\label{t:comparison}
{
	\ifx\arXiv\undefined
	\footnotesize
	\else
	\if\arXiv1
	\footnotesize
	\fi
	\fi
	\begin{tabularx}{\linewidth}{clcc} 
		\toprule
		&& SDP (interior-point solvers) & Timestepping\\
		\midrule
		&Core algorithms known in fluid mechanics&  & \Checkmark\\
		&Scalability to highly turbulent regimes &  & \Checkmark \\
		&Good practical convergence & \Checkmark & \Checkmark \\
		&Good theoretical convergence & \Checkmark &  \\
		&Easy inclusion of additional constraints & \Checkmark &  \\
		\bottomrule
	\end{tabularx}}
\end{table}

Given the poor scalability of current interior-point SDP solvers and the possibly slow convergence of the timestepping approach in dynamically complicated flow regimes (see \cref{remark:small-timesteps}),
further progress is needed before complex flows of industrial relevance can be tackled robustly and at a reasonable computational cost. We expect that significant efficiency gains may be achieved by exploiting sparsity in SDPs via matrix decomposition, as outlined in~\cref{ss:sdp-method}, as well as the newly established connection between the timestepping approach of~\cref{ss:timestepping} and low-rank factorization strategies for SDPs. Particular open questions that warrant deeper investigation include whether the (restrictive) technical assumptions required by chordal matrix decomposition can be weakened, and whether sophisticated analysis techniques and optimization algorithms for SDPs with low-rank solutions can guarantee fast convergence to the optimal bounds whilst provably avoiding spurious solutions.

Despite the many recent advances, therefore, 
the numerical analysis of the background method remains in our opinion far from complete and offers numerous opportunities for interesting research. Pursuing these opportunities promises to extend the range of flows for which bounds on mean quantities can be computed efficiently. Moreover, and perhaps more excitingly, any lessons learnt in the process may reveal practical ways to optimize  auxiliary functions more general than the quadratic ones underpinning the background method. These can in principle produce arbitrarily sharp bounds on the properties of turbulent flows~\cite{Rosa2020}, but currently remain beyond reach.

\setstretch{0.85}
\vspace{15pt}
\noindent{\small
\textbf{Acknowledgements.} We are indebted to Charlie Doering for encouraging our numerical investigation of the background method with his characteristic enthusiasm, generously hosting AW in 2013 at the University of Michigan and through fruitful discussions with GF at the Woods Hole Oceanographic Institution in 2015, 2017 and 2018. GF is also grateful to Antonis Papachristodoulou and Paul Goulart for the introduction to the world of structured SDPs, without which the connections made in \cref{ss:links} would not have been possible. Finally, we thank two anonymous reviewers and the Associate Editor for providing comments that helped us improve the original manuscript. GF was supported by an Imperial College Research Fellowship. AA is funded by the EPSRC Centre for Doctoral Training in Fluid Dynamics across Scales (award number EP/L016230/1).}

\bibliographystyle{./bibliography/imperial-vancouver}
\bibliography{./bibliography/references}

\end{document}